\documentclass[twocolumn,prd,superscriptaddress,nofootinbib,preprintnumbers]{revtex4-1}

\usepackage{mathtools,slashed}
\usepackage{booktabs,multirow}
\usepackage{graphicx}
\usepackage{amsmath,bm,amssymb,amsfonts,dsfont,xspace}
\usepackage{textgreek}
\usepackage{derivative}
\usepackage[usenames,dvipsnames]{xcolor}
\usepackage[normalem]{ulem}
\usepackage{url}
\usepackage{footmisc}
\usepackage[colorlinks  = true,
            linkcolor   = NavyBlue,
            urlcolor    = NavyBlue,
            citecolor   = NavyBlue,
            anchorcolor = NavyBlue, unicode]{hyperref}

\newcommand{\lsim}{\mathrel{\mathop{\kern 0pt \rlap
  {\raise.2ex\hbox{$<$}}}
  \lower.9ex\hbox{\kern-.190em $\approx$}}}
\newcommand{\gsim}{\mathrel{\mathop{\kern 0pt \rlap
  {\raise.2ex\hbox{$>$}}}
  \lower.9ex\hbox{\kern-.190em $\approx$}}}

\renewcommand{\vec}[1]{\boldsymbol{#1}}

\newcommand{\maddm}{{\sc MadDM}\xspace}

\newcommand{\micromegas}{{\sc micrOMEGAs}\xspace}

\usepackage[normalem]{ulem} %temporary

\usepackage{pifont}

\begin{document}

\title{WIMP Shadows: Phenomenology of Secluded Dark Matter in Three Minimal BSM Scenarios}
%Eluding the Net, Too Quiet to Catch, Whispers in the Dark, Kinetically Shy, Chemically Busy, WIMP Shadows
%Has Nature Been Playing With Us? WIMP Shadows and the Secluded Frontier
%Nature’s Quiet Joke: Why Secluded Dark Matter Eludes Detection
%Eluding the Net: Why Secluded Dark Matter Evades Direct and Indirect Searches

\author{Mattia Di Mauro}\email{dimauro.mattia@gmail.com}
\affiliation{Istituto Nazionale di Fisica Nucleare, Sezione di Torino, Via P. Giuria 1, 10125 Torino, Italy}

\author{Yanhan Wang}\email{yanhan\_wang@brown.edu}
\affiliation{Department of Physics, Brown University, Providence, RI, 02912, USA}

\date{\today}

\begin{abstract}
We present a comprehensive study of \emph{secluded} dark matter (DM) $\chi$, where the relic abundance is set by annihilations into lighter dark mediators $\phi$ that couple only feebly to the Standard Model (SM). In contrast to canonical WIMPs, which are now strongly constrained by direct and indirect searches, secluded models still achieve the observed relic abundance via thermal freeze-out into hidden-sector mediators, while predicting highly suppressed present-day signals. We analyze three minimal models: (i) a $U(1)_X$ gauge boson ($A'$) with kinetic mixing; (ii) a scalar DM candidate $S$ with a scalar mediator $K$ that has a trilinear vertex; and (iii) a Dirac fermion $\chi$ whose mass arises from a Higgs–mixed singlet $H_p$. For each model we derive annihilation and scattering rates in both WIMP-like and secluded regimes, and solve the Boltzmann equations: a single-species equation for the WIMP case and a coupled $\chi$-$\phi$ system for the secluded case to account for possible early departure of the mediator from thermal equilibrium with the SM bath. In this regard, we provide explicit lower limits on the portal coupling $\epsilon$ required to keep the mediator in thermal equilibrium with the SM bath and to ensure mediator decay before BBN. We show that for portal couplings $\epsilon\!\ll\!10^{-3}$ the relic density is dominantly controlled by DM annihilation into mediator pairs, while spin–independent scattering lies well below current limits and remains viable even for future experiments approaching the irreducible neutrino background floor. Indirect constraints are typically weak due to $p$–wave suppression, off–resonance $s$–channels, and cascade spectra controlled by $\epsilon^2$. Finally, we highlight the most promising collider tests, which remain sensitive despite tiny portal couplings.
\end{abstract}

\maketitle

\section{Introduction}\label{sec:intro}

The fundamental nature of dark matter (DM) remains one of the most profound open questions in modern physics. To date, DM has revealed itself only through its gravitational effects on astrophysical and cosmological scales, while no direct signal of DM–particle interactions has been observed in laboratory experiments~\cite{Bertone:2010zza,Bertone:2016nfn,Cirelli:2024ssz}.  
If DM is made of particles, its existence requires physics beyond the Standard Model (SM), since no known SM state can account for the observed relic abundance~\cite{Planck:2018vyg}.

A realistic DM candidate must satisfy several properties.  
It should be stable on cosmological timescales (lifetime $>$ age of the Universe), carry no electric charge, and interact very weakly with visible matter.  
The underlying BSM framework must provide a mechanism linking early-Universe dynamics to the present-day abundance.  
To match structure formation, DM must be non-relativistic by matter–radiation equality ($z\simeq 3400$) and have a sufficiently small free-streaming length (i.e.\ behaves effectively as cold dark matter).  
Moreover, its self-interaction rate must remain small enough to comply with astrophysical constraints such as those inferred from the Bullet Cluster~\cite{Clowe:2006eq}.  
Finally, any viable DM particle must satisfy the tight bounds from direct, indirect, and collider searches~\cite{Bertone:2004pz,Cirelli:2024ssz}.

Among the many candidates, Weakly Interacting Massive Particles (WIMPs) are particularly well motivated, as they can satisfy all of the above requirements~\cite{Lee:1977ua,1978ApJ...223.1015G}.  
Well-studied BSM frameworks, such as supersymmetry~\cite{WESS197439}, naturally predict new particles with the correct quantum numbers to act as WIMPs~\cite{1983PhRvL..50.1419G,Ellis:1983ew}. 
A key feature of these scenarios is that WIMPs undergo thermal freeze-out in the early Universe, which can produce the observed DM abundance.  
Remarkably, particles with electroweak-scale masses (GeV–TeV) and couplings around the weak scale naturally yield a relic density close to the measured value ($\Omega_{\rm DM} h^2 \simeq 0.12$).  
This coincidence is often referred to as the ``WIMP miracle''~\cite{Steigman:1984ac} and has driven decades of experimental efforts to detect them.

Experiments to detect WIMPs exploit three complementary strategies.  
Direct detection aims to measure nuclear or electron recoils from scattering with Galactic-halo DM particles, using large underground noble-liquid targets to reduce cosmic-ray backgrounds~\cite{Schumann:2019eaa}.  
Collider searches, performed e.g.\ at LEP and the LHC, attempt to produce DM in high-energy collisions of SM states, relying on signatures with large missing transverse energy accompanied by visible objects~\cite{Boveia:2018yeb}.  
Indirect detection looks for excesses in cosmic radiation beyond known astrophysical fluxes that may originate from DM annihilation or decay, focusing on $\gamma$ rays, neutrinos, and antimatter~\cite{Gaskins:2016cha}.  
In addition, cosmological observations, most notably measurements of the CMB, tightly constrain the DM abundance; Planck reports $\Omega_{\rm DM} h^2 = 0.120$ with $\sim\!1\%$ precision~\cite{Aghanim:2018eyx}.

Current direct-detection experiments, such as LZ and XENONnT, have reached remarkable sensitivities to spin-independent WIMP–nucleon scattering, excluding cross sections down to $\mathcal{O}(10^{-47}\text{--}10^{-48})\,\mathrm{cm}^2$ for weak-scale masses~\cite{LZ:2023,Aprile:2023XENONnT,LZ:2024zvo}.
In many canonical WIMP scenarios where the same portal controls both annihilation (setting $\Omega_{\rm DM} h^2$) and scattering, these upper limits severely restrict the viable parameter space (see, e.g.,~\cite{Arcadi:2017kky,Arcadi:2019lka,DiMauro:2023tho,Arcadi:2024ukq,DiMauro:2025jia}).  
Looking ahead, experiments such as DARWIN aim to approach the neutrino-floor sensitivity~\cite{DARWIN:2016hyl}, probing or excluding most of the remaining spin-independent WIMP parameter space.

One generic way to ease these bounds is the \emph{resonant} regime, where $m_{\rm DM}\simeq m_{\phi}/2$.  
In this kinematic configuration the $s$-channel propagator enhances the annihilation cross section, so the relic abundance can be obtained with couplings $\ll 1$, suppressing the predicted direct-detection rate.  
As a result, resonant DM models are more easily compatible with current and future limits (see, e.g.,~\cite{DiMauro:2023tho,DiMauro:2025jia}).  
That said, this scenario typically entails some degree of mass tuning unless a model-building rationale relates $m_{\phi}$ and $m_{\rm DM}$.

Alternatively, one may retain a thermal history for DM but decouple the processes setting the relic density from present-day DM–nucleus interactions.  
A simple way to reconcile a thermal relic with very weak couplings to the visible sector is the \emph{secluded} DM mechanism~\cite{Pospelov:2007mp,Pospelov:2008jd}.  
We briefly overview the main features here and apply them to three specific BSM realizations in the next section.

The dark sector contains, besides the DM particle $\chi$ with mass $m_\chi$, a lighter metastable mediator $\phi$ with mass $m_\phi$ that communicates with the SM ($\phi \phi \to \rm{SM} \,\rm{SM}$) through a small portal coupling $\epsilon\ll1$.  
In the early Universe, the dominant annihilation channel that depletes the DM density is
\begin{equation}
\chi\,\chi \;\to\; \phi\,\phi \qquad (m_\chi > m_\phi),
\end{equation}
while the mediator subsequently decays to SM states via the tiny portal.  
Thermal freeze-out then proceeds as in the canonical WIMP picture, but the relic density is set by the \emph{dark} coupling $g_X$ that controls $\chi\chi\to\phi\phi$, rather than by the portal to the SM.  
Schematically, for perturbative $s$-wave annihilation one finds
\begin{equation}
\langle \sigma v_{\rm{rel}} \rangle_{\chi\chi\to\phi\phi} \;\sim\; 
\frac{g_X^4}{16\pi\,m_\chi^2}\,
\sqrt{1-\frac{m_\phi^2}{m_\chi^2}}
\;\sim\; \mathcal{O}(10^{-26})\ \mathrm{cm^3\,s^{-1}},
\label{eq:secluded_xsec}
\end{equation}
with the precise prefactors depending on spins and interaction structures.  
Typically $g_X=\mathcal{O}(0.1\text{--}1)$, as in the WIMP paradigm and the value of the $\langle \sigma v_{\rm{rel}} \rangle$ reported in Eq.~\ref{eq:secluded_xsec} is consistent with the thermal cross section.
The portal controls only the mediator decay, $\phi\to{\rm SM\,SM}$, so DM can be \emph{thermally produced} yet effectively \emph{decoupled} from direct, indirect, and collider probes.

Two kinematic regimes are qualitatively distinct:
\begin{itemize}
\item \textbf{WIMP regime:} If $m_\chi < m_\phi$, the annihilation into $\phi\phi$ is kinematically closed and the relic abundance is set by $\chi\chi\to{\rm SM\,SM}$ through the portal coupling $\epsilon$, which takes values of the order of $\mathcal{O}(0.1-1)$ in order to achieve $\Omega_{\rm{DM}}h^2=0.12$. These values of the portal couplings are tightly constrained, and often excluded, by direct, indirect, and collider searches in the GeV–TeV mass range.

\item \textbf{Secluded regime:} If $m_\chi > m_\phi$ and the portal coupling is small, the relic density is governed by annihilation into mediator pairs (controlled by $g_X$). The mediator must still decay before Big Bang Nucleosynthesis (BBN) to avoid late energy injection. Requiring $\tau_\phi\lesssim 1$~s (i.e.\ decay before $T\sim\mathrm{MeV}$) safely avoids BBN constraints~\cite{Kawasaki:2017bqm}. In many realizations the corresponding mixing can be as low as $10^{-9}$–$10^{-11}$ and still satisfy $\tau_\phi<1$~s, depending on $m_\phi$ and the open channels (see Sec.~\ref{sec:BBN}).
\end{itemize}
Thus, for $m_\chi>m_\phi$ the thermal relic condition and experimental constraints largely decouple: $\Omega_{\rm DM}h^2$ fixes $g_X$, while direct, indirect, and collider rates are suppressed by the tiny portal $\epsilon\ll1$.

Although the portal coupling is typically very small (to satisfy direct detection and collider limits), this does not suppress the primary annihilation process $\chi\chi \to \phi\phi$, which depends only on dark-sector couplings.  
The mediators subsequently decay to SM states provided that $\Gamma_\phi^{-1}\!\ll 1$~s in order to satisfy the BBN observations. These $\phi$ decays into SM particles could be relevant for indirect-detection signals that arise as \emph{cascade} final states
\begin{equation}
\chi\chi \;\to\; \phi\phi \;\to\; 4f,\;2f+2V,\;\ldots
\end{equation}
with spectra softer than those from direct annihilation into SM particles. Moreover, the mediator decays to SM states with a width $\propto \epsilon^2$, which can reduce tension with $\gamma$-ray bounds from dwarf spheroidal galaxies (see, e.g.,~\cite{McDaniel:2023bju}).  
CMB limits on $s$-wave annihilation still apply through the effective efficiency factor $f_{\rm eff}$ for the cascade~\cite{Su:2025mxv,Xu:2024vdn}.
In Ref.~\cite{Pospelov:2008jd} the authors have investigated effects such as the Sommerfeld enhancement and bound-state formation that could significantly enhance the DM indirect detection signal. We will briefly discuss this for the first model in Sec.~\ref{sec:SEBS}.

Secluded DM thus offers a generic, economical mechanism in which a thermal relic remains hidden from direct, indirect, and collider experiments by annihilating predominantly into lighter, metastable mediators.  
If $m_\chi>m_\phi$ and $\epsilon \ll 1$, the relic density is set by dark-sector dynamics (see Eq.~\eqref{eq:secluded_xsec}), while the mediator’s tiny portal coupling only needs to satisfy $\tau_\phi\!\lesssim\!1$~s.  
This separation of roles naturally yields models in which collider, direct, and indirect rates are suppressed by many orders of magnitude with respect to the WIMP case.

In the remainder of the paper we develop a comprehensive study of secluded freeze–out across three minimal realizations, featuring both fermionic and scalar DM and scalar/vector mediators. We contrast in detail the {\it WIMP} and {\it secluded} regimes, solving the Boltzmann equation with explicit checks of kinetic and chemical equilibrium. In particular, we verify that the mediator $\phi$ in the secluded regime remains in thermal equilibrium with the SM bath by solving coupled Boltzmann equations for $\chi$ and $\phi$ and find the portal coupling values for which $\phi$ decays prior to BBN and therefore does not spoil light–element abundances.
%On the phenomenology side, we confront each model with the most up–to–date constraints: spin–independent direct–detection limits from LZ~\cite{LZ:2024zvo} and indirect–detection bounds on $\langle \sigma v_{\rm{rel}} \rangle$ from \emph{Fermi}–LAT dwarf spheroidal galaxies~\cite{McDaniel:2023bju}. The resulting outcome cleanly delineate where thermal relics survive present bounds either via resonant DM ($m_{\chi}\sim m_{\phi}/2$) or genuinely secluded dynamics and identify the portal coupling values required by thermal equilibrium of the mediator and its decay into SM particles before BBN.
On the phenomenology side, we confront each model with the latest constraints---spin-independent limits from LZ \cite{LZ:2024zvo} and \emph{Fermi}-LAT \(\gamma\)-ray bounds from dwarf spheroidal galaxies \cite{McDaniel:2023bju}---and delineate where thermal relics survive via resonant funnels or genuinely secluded dynamics. We also determine the portal values that ensure the mediator stays in thermal contact and decays before BBN.

The paper is organized as follows: Sec.~\ref{sec:models} defines the details of the three considered models; Sec.~\ref{sec:freezeout} reviews WIMP and secluded freeze--out with single and coupled Boltzmann equations and presents the BBN constraints; Sec.~\ref{sec:results} reports the combined results for the model parameters that satisfy the relic density compared to the bounds from direct- and indirect-detection; Sec.~\ref{sec:tests} includes a qualitative overview of the possible tests at collider and cosmological constraints for the model and finally in Sec.~\ref{sec:conclusion} we draw our conclusions.

\section{Models}
\label{sec:models}

In this paper, we consider three minimal BSM DM scenarios.
Throughout, $m_\chi$ denotes the fermionic DM mass, $m_S$ a complex scalar DM mass, $m_K$ and $m_{H_p}$ scalar mediators, and $m_{A'}$ vector-boson mass, sometimes called dark photon. 
%The thermal annihilation cross section is $\langle \sigma v_{\rm{rel}} \rangle_{\rm th}\simeq 2.2\times10^{-26}\,\cm^3/\s$ \cite{Planck2018}.

\subsection{Model I: $U(1)_X$ with Dirac DM, dark photon and kinetic mixing}

The first model we consider extends the SM with a new $U(1)_{X}$ group that contains a Dirac DM fermion $\chi$, which is a singlet under SM gauge symmetries, and a massive vector boson $Z'$. The Lagrangian added to the SM is
\begin{equation}
\mathcal{L} \!\supset\! \bar\chi(i\slashed{D} - m_\chi)\chi \!-\! \frac14 F'_{\mu\nu}F'^{\mu\nu} \!+\! \frac12 m_{Z'}^2 Z'_\mu Z'^\mu \!+\! \frac{\epsilon}{2}F'_{\mu\nu}B^{\mu\nu},
\label{eq:lagrU1}
\end{equation}
where the covariant derivative is $D_\mu=\partial_\mu+i q_X g_X Z'_\mu$, and $\alpha_X=g_X^2/(4\pi)$ is the dark fine-structure constant with coupling $g_X$ and charge $q_X$.
$B_{\mu\nu}$ is the field-strength tensor of the hypercharge gauge boson $B_\mu$, while
$F'_{\mu\nu}$, $g_X$, and $m_{Z'}$ are the field-strength tensor, coupling, and mass term of the new $U(1)_X$ gauge boson $Z'_{\mu}$, respectively.
The model contains a kinetic-mixing term between $B_{\mu}$ and $Z'_{\mu}$ proportional to the portal coupling $\epsilon$~\cite{Holdom:1985ag,Essig:2013lka}.
SM fermions are not charged under $U(1)_X$, while the DM fermion has $q^\chi_X=1$.
Hence SM fermions acquire an effective coupling to $Z'_{\mu}$ through kinetic mixing proportional to $\epsilon$.

The fields $B_\mu$ and $Z'_\mu$ in Eq.~\ref{eq:lagrU1} are not canonically normalized, as the kinetic terms are non-diagonal. Furthermore, they are not mass eigenstates.
After a proper rotation of the fields (see Ref.~\cite{Koechler:2025ryv} for the full calculation), we obtain the mass eigenstates associated with the photon $A_{\mu}$, the neutral electroweak boson $Z_{\mu}$, and the new gauge boson $A'_{\mu}$ with mass $m_{A'}$. 
Therefore, throughout the paper, $A'$ denotes the dark photon mass eigenstate with mass $m_{A'}$, $Z'$ is the gauge-eigenstate before diagonalization.
Assuming small kinetic mixing $\epsilon\ll 1$, the interaction Lagrangian becomes
\begin{eqnarray}
\label{eq:intL}
&&\mathcal{L}_{\text{int}} \approx - e A_\mu J_{\rm EM}^\mu - Z_\mu [g_Z \;J_Z^\mu + g_X \sin\xi  J_X^\mu] \\
&-& A'_\mu \Big[g_X\,  J_X^\mu \;-\; e \epsilon \cos\theta_W  J_{\rm EM}^\mu  + g_Z( \epsilon \sin \theta_W  -  \sin\xi )J_Z^\mu \Big]. \nonumber
\end{eqnarray}
Here, $\xi \simeq \frac{\epsilon\sin\theta_W}{1-\delta}$, $\delta \equiv m_{A'}^2/m_{Z}^2$, $\theta_W$ is the Weinberg angle, $g_Z = g/\cos{\theta_W}$ with $g$ the electroweak coupling constant, $J_{\rm{EM}}$ is the electromagnetic current, $J_Z$ is the electroweak neutral current, and $J_X$ is the $U(1)_{X}$ current defined as $J^{\mu}_X = q^\chi_X \bar{\chi} \gamma^{\mu}\chi$.
For $m_{A'} \ll m_Z$ the mixing between $A'$ and $J_Z$ becomes negligible ($\epsilon \sin \theta_W \approx \sin \xi$).
Since the new boson $A'$ can decay into quarks and leptons through kinetic mixing (see the second and third terms in the second line of Eq.~\ref{eq:intL}), the coupling $\epsilon$ is probed by collider experiments, which place strong upper limits of $\epsilon \lesssim 10^{-3}$ for $m_{A'} \in [0.1,70]~\mathrm{GeV}$ (see, e.g.,~\cite{LHCb:2017trq,Bauer:2018onh,Koechler:2025ryv}).

The annihilation channels most relevant for this model are the following.
In the \emph{WIMP regime}, when $m_\chi<m_{A'}$, the annihilation into SM fermion pairs $\chi\bar\chi\to f\bar{f}$ via off-shell $A'$ dominates and drives the relic density. A convenient form for the thermally averaged cross section is, in the non-relativistic limit \(s\simeq4 m_\chi^2\,(1+v_{\rm{rel}}^2/4)\), where $v_{\rm{rel}}$ is the DM relative velocity,
\begin{equation}
\langle \sigma v_{\rm{rel}} \rangle_{f\bar{f}} \sim \frac{N^f_c\,g_X^2 g_f^2}{2\pi}\;
\frac{m_\chi^2 \sqrt{1-\frac{m_f^2}{m_\chi^2}}\;
\left(1+\frac{m_f^2}{2 m_\chi^2}\right)}{(s - m_{A'}^2)^2 + m_{A'}^2 \Gamma_{A'}^2},
\label{eq:annU1}
\end{equation}
where $g_f$ is the effective $A'-f\bar{f}$ coupling defined as $g_f= e \, q_f\,\epsilon \cos{\theta_W}$, $q_f$ is the SM fermion electromagnetic charge, $N^f_c$ is the number of colors for the fermion $f$, and $\Gamma_{A'}$ is the $A'$ decay width.
The cross section is $s$-wave and increases significantly when $2 m_\chi \sim m_{A'}$, i.e.\ in the resonance region. The expression in Eq.~\ref{eq:annU1} scales as $\epsilon^2 g_X^2$.
Current collider constraints are of the order $\epsilon \lesssim 10^{-3}$ for $m_{A'}<70~\mathrm{GeV}$~\cite{LHCb:2017trq}. Such small values of $\epsilon$ do not allow the DM to reach the relic density with $g_X\sim 1$ away from resonance. Therefore, taking values of $m_{A'}>70~\mathrm{GeV}$ helps evade collider upper limits on $\epsilon$.

In the \emph{secluded regime} instead ($m_\chi>m_{A'}$), the relic density is dominated by the process $\chi\bar\chi\to A'A'$ occurring via $t/u$-channel exchange, which gives
\begin{eqnarray}
&& \langle \sigma v_{\rm{rel}} \rangle_{A'A'} \simeq \frac{g_X^4}{16\pi\,m_\chi^2}\;
\frac{\big(1-\frac{m_{A'}^2}{m_\chi^2}\big)^{3/2}}{\big(1-\frac{m_{A'}^2}{2m_\chi^2}\big)^2},
\label{eq:mod1sec}
\end{eqnarray}
up to $\mathcal{O}(1)$ factors away from threshold.
The channel \(\chi\bar\chi\to A'A'\) is \(s\)-wave for Dirac \(\chi\) with vector couplings; for Majorana \(\chi\) the leading term would be \(p\)-wave.
The annihilation into $ZZ$, which has the same expression as Eq.~\ref{eq:mod1sec} with an additional factor $\sin^4\xi$ and the replacement $m_{A'}\to m_Z$, has the same strength as the annihilation into $A'A'$ only when $\epsilon\sim \mathcal{O}(1)$.

The nuclear cross section relevant for direct detection arises from $t$-channel exchange of $A'$ and $Z$ between DM and quarks.
We first define the quark–level couplings
\begin{equation}
C_u^{M}(q^2)\equiv \frac{g_{\chi M}\,g_{M u}^{(V)}}{m_{M}^{2}+q^{2}},
\qquad
C_d^{M}(q^2)\equiv \frac{g_{\chi M}\,g_{M d}^{(V)}}{m_{M}^{2}+q^{2}},
\end{equation}
where \(g_{\chi M}\) is the \(\chi\)–\(M\) vector coupling and \(g_{M q}^{(V)}\) the vector coupling of \(M\) to quark \(q\) (with \(M=A',Z\)).
The coherent proton/neutron couplings are
\begin{eqnarray}
f_p(q^2)&=&\sum_{M=A',Z}\big[\,2C_u^{M}(q^2)+C_d^{M}(q^2) \big], \\
f_n(q^2)&=&\sum_{M=A',Z}\big[\,C_u^{M}(q^2)+2C_d^{M}(q^2) \big].
\end{eqnarray}
For a nucleus with \((A,Z)\) and form factor \(F(q)^2\) the nuclear cross section is
\begin{equation}
\sigma_{\chi N}^{\rm SI}(q) \;=\;
\frac{\mu_{\chi N}^2}{\pi}\;
\Big[\, Z f_p(q^2) + (A-Z) f_n(q^2) \,\Big]^2\; F(q)^2,
\label{eq:sigmaA_SI}
\end{equation}
where \(\mu_{\chi N}\) is the \(\chi\)–nucleus reduced mass.
%The complete expression of $f_p(q^2)$ and $f_n(q^2)$ can be found, for example, in Ref.~\cite{DEramo:2016gos}.

Direct-detection experiments typically operate in the zero–momentum-transfer limit ($q \ll m_{A'}$).
%, under which the spin–independent cross section on a nucleon $N=p,n$ reduces to
%\begin{eqnarray}
%\sigma_{\chi N}^{\rm SI}
%&=&\frac{\mu_{\chi N}^2}{\pi}
%\left\{\!
%\frac{g_{\chi A'}}{m_{A'}^2}
%\Big[\,A\big(g_{A'u}^{(V)} \! + \! 2g_{A'd}^{(V)}\big)
%+ Z\big(g_{A'u}^{(V)}-g_{A'd}^{(V)}\big)\Big] \! \right\} +
%\nonumber\\
%&+& \frac{\mu_{\chi N}^2}{\pi}\left\{\frac{g_{\chi Z}\,g_Z}{m_Z^2}\!
%\left[\,\frac{Z-A}{2}-2s_W^2 Z\,\right]
%\right\}^{2}.
%\label{eq:DDU1X}
%\end{eqnarray}
In this case and in the limit $m_{A'}\ll m_Z$, relevant for the secluded regime, this simplifies to
\begin{equation}
\sigma_{\chi p}^{\rm SI}=\frac{\mu_{\chi p}^2}{\pi}\,
\frac{(g_X q_X\,\epsilon e \cos\theta_W)^2}{m_{A'}^4},
\qquad
\sigma_{\chi n}^{\rm SI}=0\,,
\label{sec:mod1DD}
\end{equation}
so scattering is maximally isospin-violating (proton-only).

%The nuclear cross section for direct detection is spin independent and For a heavy mediator,
%\begin{align}
%\sigma_{\chi N}^{\rm SI}\simeq \frac{16\pi\,\alpha\,\alpha_D\,\epsilon^2\,\mu_{\chi N}^2}{m_{Z'}^4}.
%\end{align}
%Model I has been widely used in dark--sector and dark--photon studies \cite{Essig:2013lka,Alexander:2016aln,Battaglieri:2017aum}.

\subsection{Model II: {\tt DMSimp} with scalar $S$ DM and scalar mediator $K$}
\label{sec:modelscalar}

In the second model we consider the addition to the SM of a scalar mediator $K$ and a scalar DM particle $S$, which have masses $m_K$ and $m_S$, respectively.
This model is typically studied within the DM simplified models ({\tt DMSimps}) framework (see, e.g., \cite{Arina:2018zcq,DiMauro:2025jia}).
The Lagrangian of the model is
\begin{eqnarray}
\label{eq:simplscalar}
&&\mathcal{L} \supset \partial_\mu K \partial^\mu K - m_K^2|K|^2 + \frac12 \partial_\mu S \partial^\mu S - \frac12 m_S^2 S^2 \\
&+&  m_{K} g_{K} K^3 + m_S \lambda_{KS}S^2 K + \lambda_{KS}^2S^2K^2 + \frac{\beta}{\sqrt{2}} \frac{m_f}{v_h} \bar{f} f K. \nonumber
\end{eqnarray}
In the first line of the Lagrangian we report the kinetic and mass terms for $S$ and $K$.
The second line contains a cubic term for $K$, the decay term of the mediator into a DM pair, a point-like annihilation term between the DM and the mediator, and finally the coupling between the mediator and fermions, which is essentially a Yukawa coupling proportional to the parameter $\beta$ and the fermion mass.
Optionally $S$ may (weakly) mix with the Higgs via $\mu S H^\dagger H + \lambda_{HS}S^2H^\dagger H$ \cite{Silveira:1985rk,McDonald:1993ex,Burgess:2000yq}. However, we do not take this into account for this model; it will be considered in Sec.~\ref{sec:higgs}.

In the WIMP regime $m_S<m_K$, the most relevant annihilation cross section for the relic density is the one producing fermions, whose expression in the non-relativistic limit \(s\simeq 4 m_S^2\,(1+v_{\rm rel}^2/4)\) is
\begin{equation}
\langle \sigma v_{\rm{rel}} \rangle_{f\bar f} \approx
\frac{N_c^f m_S^2 \lambda_{KS}^2 \beta m_f^2}{16\pi s \, v_h^2}
\frac{\left(1-\frac{m_f^2}{m_S^2}\right)^{3/2}}{(4 m_S^2- m_K^2)^2+\Gamma_K^2 m_S^2},
\label{eq:sig_ff_exact}
\end{equation}
which is \(s\)-wave dominated and helicity suppressed (i.e.~proportional to $m_f^2/m_{S}^2$).

In the secluded regime, instead, the relic-density cross section is dominated by the annihilation process \(S S \to KK\).
The tree-level amplitude of this interaction receives contributions from: (i) the contact term \(S^2K^2\) proportional to $\lambda_{KS}^2$, (ii) \(t/u\)-channel exchange of \(S\) via \(m_S \lambda_{KS}\), and (iii) an \(s\)-channel \(K\) via \(m_S \lambda_{KS}\) and \(m_{K}g_{K}\).
With identical bosons in the final state, the non-relativistic leading term is \(s\)-wave and given by
\begin{equation}
\langle \sigma v_{\rm{rel}} \rangle_{KK}
\approx \frac{1}{64\pi\,m_S^2}\;
\sqrt{1-\frac{m_K^2}{m_S^2}}\;
\big|\mathcal{A}_0\big|^2,
\label{eq:sigv_YY_NR}
\end{equation}
where the \emph{angle–averaged} amplitude is
\begin{equation}
\label{eq:amplitude}
\mathcal{A}_0  \equiv 
\lambda_{KS}^2 - \frac{2\,m_S^2 \lambda_{KS}^2}{m_S^2-m_K^2}
+ \frac{m_S \lambda_{KS}\, m_K g_{K}}{4m_S^2-m_K^2+i m_K\Gamma_K}.
\end{equation}
The annihilation cross section into $KK$ is thus proportional to $\lambda_{KS}^4$ or $\lambda_{KS}^2 g_K^2$, depending on which term in Eq.~\ref{eq:amplitude} dominates.

The direct-detection cross section of this model is spin–independent.
Integrating out the mediator \(K\) (since $q^2\ll m_K^2$), the Lagrangian in Eq.~\ref{eq:simplscalar} yields the scalar operator $\mathcal{L}_{\rm eff}\supset m_S \lambda_{KS} \beta/(\sqrt{2} v_h m_K^2)\,S^2 \sum_q m_q\,\bar q q$.
Matching to nucleons gives
\begin{eqnarray}
g_{SNN} &=& \sum_{q=u,d,s} g_f\, f_q^N
+ \frac{2}{27}\,f_{TG}^N \sum_{Q=c,b,t} g_Q ,\\
f_{TG}^N &\equiv& 1-\sum_{q=u,d,s} f_q^N, \nonumber
\end{eqnarray}
with the usual scalar form factors \(f_q^N\) defined by $\langle N | m_q \bar q q | N \rangle \;=\; m_N f_{q}^{N}$.
The effective scalar coupling to a nucleon $N=p,n$ reads
\begin{equation}
f_N(q^2) \;=\;
\frac{m_S \lambda_{KS}}{q^2+m_K^2}\;
\frac{\beta}{\sqrt{2}}\;\frac{m_N}{v_h}\;
\bigg[ \sum_{q=u,d,s} f_{q}^{N} + \frac{2}{9} f_{TG}^{N} \bigg].
\label{eq:fN}
\end{equation}
The spin–independent per–nucleon cross section in the limit $q^2 \to 0$ becomes
\begin{equation}
\sigma^{\text{SI}}_{SN} =
\frac{\mu_N^2 m_S^2 \lambda^2_{KS} \beta^2 m_N^2}{2\pi \,m_K^4 v_h^2}
\left( \sum_{q=u,d,s} f_{Tq}^{(N)} + \frac{2}{9} f_{TG}^{(N)} \right)^2 ,
\label{eq:sigmaSI}
\end{equation}
where $\mu_N = m_S m_N / (m_S+m_N)$ is the reduced mass of the DM–nucleon system, and $\sum_{q=u,d,s} f_{Tq}^{(N)} + \frac{2}{9} f_{TG}^{(N)}\sim 0.3$.

\subsection{Model III: $U(1)_X$ with Dirac DM and a Higgs--mixed singlet generating $m_\chi$}
\label{sec:higgs}

Analogously to Model~I, we extend the SM gauge group by an extra abelian factor $U(1)_X$ with gauge boson $Z'_\mu$ and coupling $g_X$. Here the dark sector contains a Dirac fermion DM particle $\chi$ with $U(1)_X$ charge $q_X$, and a complex singlet scalar $R$ that mixes with the SM Higgs doublet $\Phi$ (see, e.g., \cite{Baek:2011aa}). We allow for kinetic mixing between the two abelian field strengths with parameter $\epsilon$ but, contrary to what was done in Model~I, we take $\epsilon$ so small that kinetic mixing plays no role in the phenomenology.

Two frequently used charge assignments are:
\begin{itemize}
\item \textbf{Vector-like $\chi$ under $U(1)_X$:} the left- and right-handed components of the DM particle have the same charge $q_{\chi_L}=q_{\chi_R}=q_\chi$. A bare Dirac mass $m_0\,\bar\chi\chi$ is gauge invariant, and a renormalizable Yukawa portal $-\;y_p\,\bar\chi\chi R$ requires the complex scalar $R$ to be neutral under $U(1)_X$ ($q_R=0$). In this case $U(1)_X$ remains unbroken and the $Z'$ boson cannot acquire mass via a Higgs-like mechanism\footnote{The kinetic term $(D_\mu R)^\dagger (D^\mu R)$ is typically the one that, through the Brout–Englert–Higgs mechanism, provides mass to a gauge boson. However, since $q_R=0$, $D_\mu R=\partial_\mu R$ and no mass term for $Z'$ emerges from the $R$ kinetic term.}.
\item \textbf{Chiral $\chi$ under $U(1)_X$:} the two $\chi$ spinors carry different charges, $q_{\chi_L}\neq q_{\chi_R}$. The bare mass term is therefore forbidden (as for SM fermions) and a gauge-invariant Yukawa $-\;y_p\,\bar\chi_L\chi_R R+\text{h.c.}$ requires $q_R=q_{\chi_L}-q_{\chi_R}$. Here $R$ can also break $U(1)_X$ and give a Higgs-like mass to $Z'$. In particular, $Z'$ acquires a mass $M_{Z'}=g_X |q_R| v_R$ and the CP-odd phase of $R$ is the Goldstone eaten by $Z'$.
\end{itemize}
In the rest of the paper we consider the vector-like $\chi$ case where $Z'$ may acquire a Stückelberg mass (no charged scalar needed)\footnote{%
An alternative to the Higgs mechanism is that the new gauge boson $Z'_\mu$ acquires its mass via the Stückelberg mechanism. 
Introduce a pseudoscalar axion-like field $a$ that shifts under $U(1)_X$ as $a \to a + m\,\alpha(x)$ when $Z'_\mu\to Z'_\mu - \partial_\mu \alpha(x)/g_X$. 
The gauge-invariant Stückelberg Lagrangian reads
$\mathcal{L}_{\rm St}=-\frac{1}{4} F'_{\mu\nu}F'^{\mu\nu}
+ \frac{1}{2}\,\big(m_{Z'} Z'_\mu + \partial_\mu a \big)^2$,
where $m_{Z'}$ is the Stückelberg mass parameter.
After gauge fixing, $a$ is eaten as the longitudinal mode of $Z'$, which thereby acquires the mass $m_{Z'}$ without a scalar VEV.
The Stückelberg construction is renormalizable and consistent at the EFT level; see, e.g., \cite{Kors:2004dx, Feldman:2007wj, Ruegg:2003ps}.}.

The relevant pieces of the model are
\begin{eqnarray}
\mathcal{L} &\supset&
-\frac{1}{4} F'_{\mu\nu}F'^{\mu\nu}
-\frac{\epsilon}{2}\, F'_{\mu\nu} B^{\mu\nu}
+ (D_\mu R)^\dagger (D^\mu R)
- V(\Phi,R)
\nonumber\\
&\quad&
+\bar\chi\,(i\!\not\!\! D - m_\chi)\,\chi
-\Big( y_p\,\bar\chi\chi\,R + \text{h.c.}\Big),
\label{eq:Ltot}
\end{eqnarray}
with the covariant derivative for the new states defined as $D_\mu=\partial_\mu - i g_X q_X\, Z'_\mu$, and the potential
\begin{eqnarray}
V(\Phi,R)
&=& \mu_H^2\,\Phi^\dagger\Phi + \mu_R^2\, R^\dagger R
+\lambda_H (\Phi^\dagger\Phi)^2 \\
&+& \lambda_R (R^\dagger R)^2
+\lambda_{HR} (\Phi^\dagger\Phi)(R^\dagger R)\, . \nonumber
\end{eqnarray}
The term $y_p\,\bar\chi\chi\,R$ is a Yukawa interaction between $R$ and the DM particle.
A key advantage of the vector-like case is that $y_p$ can be chosen independently of the DM mass $m_\chi$. In fact, $m_\chi=m_0 + y_p v_R/\sqrt{2}$, where $m_0$ is the bare mass term. In the minimal chiral case one instead has $m_\chi=y_p v_R/\sqrt{2}$, so $m_\chi$ and $y_p$ are related.

Electroweak symmetry breaking and scalar-sector mixing are parameterized as
\begin{equation}
\Phi=\begin{pmatrix} G^+ \\ \dfrac{v_h+h+i G^0}{\sqrt{2}} \end{pmatrix},
\qquad
R=\dfrac{v_R+\rho}{\sqrt{2}}\,,
\end{equation}
with $v_h\simeq 246~\text{GeV}$ the vacuum expectation value (VEV) of the SM Higgs doublet $\Phi$, and $v_R$ the VEV of the singlet scalar $R$. 
The fields $G^+$ and $G^0$ are the would-be Goldstone bosons, absorbed by $W^\pm$ and $Z$ after electroweak symmetry breaking. No Goldstone arises from $R$ because $U(1)_X$ is unbroken ($q_R=0$).
In the CP-even basis $(h,\rho)$ the mass matrix is
\begin{equation}
\mathcal{M}^2=
\begin{pmatrix}
2\lambda_H v_h^2 & \lambda_{HR} v_h v_R \\
\lambda_{HR} v_h v_R & 2\lambda_R v_R^2
\end{pmatrix},
\,
\tan 2\alpha = \frac{\lambda_{HR} v_h v_R}{\lambda_R v_R^2 - \lambda_H v_h^2}\,,
\end{equation}
where $\alpha$ is the mixing angle. The mass eigenstates $H$ (SM-like) and $H_p$ (singlet-like) are
\begin{equation}
\begin{pmatrix} H \\ H_p \end{pmatrix}
=
\begin{pmatrix}
\cos\alpha & \sin\alpha \\
-\sin\alpha & \cos\alpha
\end{pmatrix}
\begin{pmatrix} h \\ \rho \end{pmatrix},
\end{equation}
with masses
\begin{equation}
    m_{H,H_p}^2=\lambda_H v_h^2 + \lambda_R v_R^2 \mp
\sqrt{\big(\lambda_R v_R^2-\lambda_H v_h^2\big)^2+(\lambda_{HR} v_h v_R)^2}.
\end{equation}
Couplings to fermions and DM follow from this rotation. In particular,
\begin{align}
g_{\chi\chi H} &= \frac{y_p}{\sqrt{2}}\,\sin\alpha\,, &
g_{\chi\chi H_p} &= \frac{y_p}{\sqrt{2}}\,\cos\alpha\,, \\
g_{H f\bar f} &= \frac{m_f}{v_h}\,\cos\alpha\,, &
g_{H_p f\bar f} &= -\,\frac{m_f}{v_h}\,\sin\alpha\,,
\end{align}
and $Z'$ couples to $\chi$ with $g_\chi=g_X q_\chi$.
We do not include any interaction between $Z'$ and SM fermions, either via $U(1)_X$ charges or through kinetic mixing $\epsilon$.

In the WIMP regime the dominant channels are $\chi\bar\chi\to f\bar f$ via $s$-channel $H,H_p$.
Away from resonances and in the non-relativistic limit,
\begin{widetext}
\begin{equation}
\langle \sigma v_{\rm{rel}} \rangle_{ff}
\simeq
\sum_f \frac{N_c^f\,y_p^2\,m_f^2}{2\pi v_h^2}\,
\left(1-\frac{m_f^2}{m_\chi^2}\right)^{3/2}\,
\left|
\frac{\sin\alpha\,\cos\alpha}{4 m_\chi^2-m_H^2+i m_H\Gamma_H}
-\frac{\sin\alpha\,\cos\alpha}{4 m_\chi^2-m_{H_p}^2+i m_{H_p}\Gamma_{H_p}}
\right|^2.
\label{eq:sigmaff-scalar}
\end{equation}
\end{widetext}
Equation~\eqref{eq:sigmaff-scalar} contains the sum over kinematically open annihilation channel into fermions and it is helicity suppressed by $m_f^2$ in the $s$-wave and exhibits resonant enhancement near $2m_\chi\simeq m_{H,H_p}$. $\Gamma_H$ and $\Gamma_{H_p}$ are the $H$ and $H_p$ decay widths.
Because the amplitude is proportional to $\sin\alpha\,\cos\alpha$, in the small-mixing limit the indirect-detection constraints from $\chi\bar\chi\to f\bar f$ are negligible.

In the secluded regime, if $m_\chi > m_{H_p}$ (and/or $m_\chi > m_{Z'}$), annihilations into dark-sector states dominate, e.g.~into $H_p H_p$ or $Z'Z'$:
\begin{align}
\langle \sigma v_{\rm{rel}} \rangle_{H_p H_p}
&\simeq
\frac{y_p^4 (\cos{\alpha})^4\, v_{\rm{rel}}^2}{64\pi m_\chi^2}\,
\frac{ \left(1-\frac{m_{H_p}^2}{m_\chi^2}\right)^{3/2} }{ \left(1-\frac{m_{H_p}^2}{2m_\chi^2}\right)^{4} },
\label{eq:sigmaHH}
\\
\langle \sigma v_{\rm{rel}} \rangle_{Z'Z'}
&\simeq
\frac{(g_X q_\chi)^4}{16\pi m_\chi^2}\;
\frac{\left(1-\frac{M_{Z'}^2}{m_\chi^2}\right)^{3/2}}
{\left(1-\frac{M_{Z'}^2}{2 m_\chi^2}\right)^2},
\label{eq:sigmaZZ}
\end{align}
where the former arises from $t$- and $u$-channel $\chi$ exchange and is $p$-wave suppressed, while the latter is an unsuppressed $s$-wave.

There is also an $s$-channel diagram contributing to $\chi\bar\chi\to H_p H_p$ whose non-relativistic form is
\begin{equation}
\langle \sigma v_{\rm{rel}} \rangle_{s\text{-ch}}
\simeq
\frac{\sqrt{1-\frac{m_{H_p}^{2}}{m_\chi^{2}}}}{64\pi\,m_\chi^{2}}\,
\left|
\sum_{i=H,H_p}
\frac{\,g_{\chi\chi h_i}\,\lambda_{h_i H_p H_p}\,}{\,4m_\chi^2-m_{h_i}^2+i\,m_{h_i}\Gamma_{h_i}\,}
\right|^2,
\label{eq:sigma_schannel_NR}
\end{equation}
where \(\lambda_{h_i H_p H_p}\) are the trilinear scalar couplings in the mass basis (\(h_i\in\{H,H_p\}\)).
For indirect detection this $s$-channel piece with $H_p$ as a mediator typically dominates; however, the subsequent decay $H_p\to f\bar f$ is proportional to $g_{H_p f\bar f}\propto \sin\alpha$, so for very small mixing even this channel yields weak indirect constraints.

For the relic density, the $p$-wave contribution in Eq.~\eqref{eq:sigmaHH} is not strongly suppressed because the DM relative velocity at freeze-out is $v_{\rm{rel}}\sim 0.3$. Depending on the mixing angle, the scalar masses, and $y_p$, the $t$/$u$ and $s$-channel pieces can contribute comparably to the thermally averaged cross section. As a rule of thumb, the $p$-wave term scales as $y_p^4/m_\chi^2$, the off-shell $s$-channel (for $m_\chi>m_{H_p}$, away from resonance) scales roughly as $y_p^2/m_\chi^{6}$ and the interference between the $s$ and the $t/u$ channels is proportional to $y_p^2/m_\chi^4$.

\medskip

Spin-independent scattering off a nucleon $N=p,n$ receives $t$-channel contributions from $H$ or $H_p$ exchange:
\begin{eqnarray}
\sigma^{\rm SI}_{\chi N} & =&
\frac{\mu_N^2}{\pi}\left[
\sum_{i=H,H_p}
\frac{g_{\chi\chi h_i}\, g_{h_i NN}}{m_{h_i}^2}
\right]^2, \\[4pt]
g_{h_i NN}&=&\frac{m_N}{v_h}\, f_N \times
\begin{cases}
\cos\alpha & (h_i=H),\\[2pt]
-\sin\alpha & (h_i=H_p),
\end{cases}
\label{eq:SI-scalar}
\end{eqnarray}
where $\mu_N$ is the $\chi$–$N$ reduced mass and $f_N\simeq 0.30$, as in Eq.~\ref{eq:fN}, encodes the nucleon scalar form factors.
Putting the pieces together,
\begin{equation}
\sigma^{\rm SI}_{\chi N} =
\frac{\mu_N^2\, m_N^2\, f_N^2\, \cos^{2}\!\alpha\,\sin^{2}\!\alpha\, y_p^2}{2\pi\, v_h^2 }
\left(\frac{1}{m_H^2}-\frac{1}{m_{H_p}^2}\right)^2,
\end{equation}
which exhibits a cancellation when $m_H\simeq m_{H_p}$ (``blind spot'').
%Using the couplings above,
%\begin{equation}
%g_{\chi\chi H}=\frac{y_p}{\sqrt{2}}\sin\alpha,\qquad
%g_{\chi\chi H_p}=\frac{y_p}{\sqrt{2}}\cos\alpha.
%\end{equation}
Therefore, in this model there are two conditions that can reduce significantly direct detection signal: i) $\sin{\alpha}\ll 1$ and ii) $m_H\simeq m_{H_p}$. We will investigate in this paper the first condition that is relevant for secluded DM models.

%\paragraph{Relic target.}
%The observed abundance $\Omega_\chi h^2\simeq 0.12$ is reproduced for
%$\langle \sigma v_{\rm{rel}} \rangle_{\rm fo}\sim(2{-}3)\times 10^{-26}\ \text{cm}^3\text{/s}$ in standard cosmology; Eqs.~(\ref{eq:sigmaff-scalar})–(\ref{eq:sigmaZZ}) offer several viable pathways: resonance ($2m_\chi\simeq m_{h_i}$), $Z'$ portals, or secluded annihilations into $H_pH_p$/$Z'Z'$.

\section{Relic density in the WIMP and Secluded regimes}
\label{sec:freezeout}

\subsection{Solution of the Boltzmann equation in the Standard WIMP case}

The cosmic abundance of DM today is tightly constrained by the Planck measurement $\Omega_{\rm{DM}} h^2 = 0.120 \pm 0.001$, which is at the percent-level~\cite{Planck:2018vyg}. 
Therefore, by matching the DM parameters to the observed $\Omega_{\rm{DM}} h^2$ provides one of the strongest constraints to the BSM phenomenology.

In an expanding Friedmann–Robertson-Lemaître–Walker spacetime, the phase–space distribution
$f_\chi(\vec p)$ of a DM particle $\chi$ obeys the Boltzmann equation
\begin{equation}
E\Big(\partial_t - H\,\vec p\!\cdot\!\nabla_{\vec p}\Big) f_\chi(\vec p)
= \mathcal{C}[f_\chi]\,, 
\label{eq:boltzmann}
\end{equation}
with $E$ and $\vec p$ the single–particle energy and momentum, and $H$ the Hubble rate~\cite{Kolb:1990vq,Edsjo:1997bg}. 
The collision operator $\mathcal{C}$ contains elastic terms (maintaining kinetic equilibrium with the thermal bath) and annihilation terms (controlling chemical equilibrium).

For standard WIMP freeze–out one assumes that \emph{kinetic} equilibrium holds during chemical decoupling, so that $f_\chi$ remains proportional to the equilibrium distribution. 
This is true for most of the parameter space except very close to the resonance region, i.e.~when $m_\chi$ is about half of the mediator mass (see e.g.~\cite{Binder:2017rgn} for a detailed discussion).
Under this assumption, Eq.~\eqref{eq:boltzmann} reduces to an ordinary differential Lee-Weinberg equation describing the evolution of the DM number density $n_{\chi}$
\begin{equation}
\frac{d n_\chi}{dt} + 3 H n_\chi
= - \left\langle \sigma v_{\mathrm{M\o l}} \right\rangle_T
\left(n_\chi^2 - n_{\chi,\mathrm{eq}}^{\,2}\right),
\label{eq:leeweinberg}
\end{equation}
with $n_\chi = g_\chi \int \!\frac{d^3p}{(2\pi)^3} f_\chi(p)$ and the thermal average of the M\o ller velocity–weighted cross section evaluated at temperature $T$, $\langle\sigma v_{\mathrm{M\o l}}\rangle_T$.
The Hubble rate in radiation-dominated era is given by:
% Radiation domination:
\begin{equation}
H^2(T)=\frac{8\pi G}{3}\,\rho_{\rm rad}(T), 
\,\, 
\rho_{\rm rad}(T)=\frac{\pi^2}{30}\,g_\ast(T)\,T^4 ,
\end{equation}
where $g_\ast$ is effective number of relativistic degrees of freedom that contribute to the energy density of the Universe at a given temperature.

When $\chi$ is non–relativistic ($f_{\chi,\mathrm{eq}}\!\propto e^{-E/T}$), the thermally averaged annihilation cross section reads~\cite{GONDOLO1991145}
\begin{equation}
\left\langle \sigma v_{\mathrm{M\o l}} \right\rangle_T
= \int_{4 m_\chi^2}^{\infty} \! ds\;
\frac{s\,\sqrt{s-4 m_\chi^2}\; K_1(\sqrt{s}/T)\; \sigma v_{\mathrm{M\o l}}(s)}
{16\,T\,m_\chi^4\,K_2^2(m_\chi/T)}\,,
\label{eq:thermalavg}
\end{equation}
where $K_\nu$ are modified Bessel functions of the second kind of order $\nu$.
In the annihilation cross sections, the Møller velocity $v_{\text{M\o l}}$ represents the relative velocity between the two DM particles ($v_{\rm{rel}}$), defined as $v_{\text{M\o l}} = \sqrt{(p_1\!\cdot\!p_2)^2 - m_\chi^4}/(E_1E_2)$, which in the non--relativistic limit reduces to the usual relative speed $v_{\text{rel}}\simeq|\mathbf{v}_1-\mathbf{v}_2|$.

It is convenient to rewrite the Lee--Weinberg equation in terms of the comoving abundance
\(Y \equiv n_\chi / s(T)\), where \(s(T)\) is the entropy density, and to evolve it as a function of
\(x \equiv m_\chi/T\).
Freeze-out approximately occurs at a temperature $T_f$ when the \emph{per-particle} annihilation rate $\Gamma_{\rm ann} \;\equiv\; n_\chi\,\big\langle \sigma v_{\mathrm{M\o l}}\big\rangle_T$
falls below the Hubble expansion rate \(H\).

For weak-scale WIMPs one typically finds \(x_f \equiv m_\chi/T_f \simeq 20\text{--}30\),
with \(g_\ast \sim 90\) relativistic degrees of freedom at \(T_f\).
At this point chemical decoupling sets in and the comoving abundance approaches its asymptotic value $Y_0$
\begin{equation}
Y_0 \simeq
\frac{3.79\,x_f}{\sqrt{g_\ast}\,M_{\rm Pl}\, m_\chi\,
\langle \sigma v_{\rm{rel}} \rangle}\,,
\label{eq:Y0}
\end{equation}
following the standard treatment~\cite{Kolb:1990vq,GONDOLO1991145}. 
The present–day relic abundance then becomes
\begin{equation}
\Omega_{\rm{DM}} h^{2}
\simeq
\frac{m_\chi s_0 Y_0}{\rho_{\rm c}}
\approx
1.07\times 10^{9}\;{\rm GeV}^{-1}\;
\frac{x_f}{\sqrt{g_\ast}\,M_{\rm Pl}}\;
\frac{1}{\langle \sigma v_{\rm{rel}} \rangle}\,,
\label{eq:omega}
\end{equation}
with $M_{\rm Pl}=1.22\times10^{19}\,\mathrm{GeV}$, today’s entropy density $s_0=2.89\times10^3\,\mathrm{cm}^{-3}$, and critical density $\rho_{\rm c}=8.05\times10^{-47}\,h^2\,\mathrm{GeV}^4$.
Matching the observed $\Omega_{\rm{DM}} h^2\simeq 0.12$ yields the canonical thermal cross section \cite{KolbTurner,Gondolo:1990dk}
\begin{equation}
\langle \sigma v_{\rm{rel}} \rangle \simeq
3\times 10^{-26}\,\mathrm{cm^3\,s^{-1}}\,
\left(\frac{x_f}{25}\right)\!
\left(\frac{90}{g_\ast}\right)^{1/2},
\label{eq:svcanonical}
\end{equation}
which serves as a benchmark for WIMP annihilation in the early Universe \cite{Steigman:2012nb,Bringmann:2020mgx}.
%For $s$--wave annihilation, the approximate relic abundance is
%\begin{align}
%\Omega_\chi h^2 \simeq \frac{0.12}{\langle \sigma v_{\rm{rel}} \rangle/(2.2\times10^{-26}\ \cm^3/\s)},
%\end{align}
%up to $\order(10\%)$ corrections from $x_f$ and $g_\star$ \cite{KolbTurner,Gondolo:1990dk,Planck2018}.

\subsection{Solution of the Boltzmann equation in the Secluded regime}
\label{sec:secluded}

When a dark-sector particle $\phi$ (mediator or scalar) is lighter than the DM state $\chi$, $m_\chi>m_\phi$, the channel $\chi\bar\chi\!\to\!\phi\phi$ opens and can dominate over annihilations into SM particles, $\chi\bar\chi\!\to\!\mathrm{SM}\,\mathrm{SM}$. In this \emph{secluded} regime the dark sector may chemically decouple from the SM before DM freeze--out, so one must in general track the abundances of both $\phi$ and $\chi$. In order to follow the evolution of dark-sector particles we can use the following convenient form of the coupled Lee--Weinberg system for the number densities of the DM and mediator:
\begin{widetext}
\begin{eqnarray}
\frac{dn_\chi}{dt}+3H n_\chi
&=& -\langle \sigma v_{\rm{rel}} \rangle_{\chi\bar\chi\to\phi\phi}
\!\left(n_\chi^2 - n_{\chi,{\rm eq}}^{\,2}\,\frac{n_\phi^2}{n_{\phi,{\rm eq}}^{\,2}}\right)
\;-\; \langle \sigma v_{\rm{rel}} \rangle_{\chi\bar\chi\to{\rm SM}}
\!\left(n_\chi^2 - n_{\chi,{\rm eq}}^{\,2}\right), \label{eq:BoltzPhi_chi}\\[4pt]
\frac{dn_\phi}{dt}+3H n_\phi
&=& +\langle \sigma v_{\rm{rel}} \rangle_{\chi\bar\chi\to\phi\phi}
\!\left(n_\chi^2 - n_{\chi,{\rm eq}}^{\,2}\,\frac{n_\phi^2}{n_{\phi,{\rm eq}}^{\,2}}\right)
\;-\; \langle \sigma v_{\rm{rel}} \rangle_{\phi\phi\to{\rm SM}}
\!\left(n_\phi^2 - n_{\phi,{\rm eq}}^{\,2}\right)
\;-\; \langle\Gamma_\phi\rangle\left(n_\phi-n_{\phi,{\rm eq}}\right).
\label{eq:BoltzPhi}
\end{eqnarray}
\end{widetext}

In the secluded regime the mediator $\phi$ can stay in thermal equilibrium with the SM bath, i.e.\ maintain chemical and kinetic equilibrium, by the following three types of processes:
\begin{itemize}
\item decays and inverse decays, $\phi \leftrightarrow {\rm SM}\,{\rm SM}$, with thermally averaged rate
\begin{equation}
\langle\Gamma_\phi\rangle(T) \equiv \Gamma_\phi\,\frac{K_1(m_\phi/T)}{K_2(m_\phi/T)}\,,
\end{equation}
where $\Gamma_{\phi}$ is the $\phi$ decay width at rest;

\item annihilations and inverse annihilations into SM particles, $\phi\phi \leftrightarrow {\rm SM}\,{\rm SM}$ (or $\phi\phi^\dagger$ if $\phi$ is complex), with rate
\begin{equation}
\Gamma_{\rm ann}(T) \equiv n_\phi^{\rm eq}(T)\,\langle \sigma v_{\rm{rel}} \rangle_{\phi\phi\to {\rm SM\,SM}}\,,
\end{equation}
where $n_\phi^{\rm eq}$ is the equilibrium density of $\phi$;

\item elastic scattering, $\phi f \leftrightarrow \phi f$, with rate
\begin{equation}
\Gamma_{\rm el}(T) \equiv \sum_{f} n_f^{\rm eq}(T)\,\langle \sigma v_{\rm{rel}} \rangle_{\phi f\to \phi f}\,,
\end{equation}
which maintains kinetic equilibrium but does not change $n_\phi$.
\end{itemize}

The $2\!\to\!2$ \emph{elastic} scattering of $\phi$ with the bath (e.g.\ $\phi f\!\leftrightarrow\!\phi f$) maintains \emph{kinetic} equilibrium provided
\begin{equation}
\Gamma_{\rm el}(T)\;\equiv\;\sum_f n_f^{\rm eq}(T)\,\langle \sigma v_{\rm{rel}} \rangle_{\phi f\to \phi f}\ \gg\ H(T)\,.
\end{equation}
If the kinetic equilibrium of $\phi$ is maintained, its momentum distribution matches a thermal shape with the bath temperature $T$ and the hidden sector shares the bath temperature, $T' = T$. Elastic scattering enters only the \emph{kinetic} sector and it does not change the DM number density; thus the term $\Gamma_{\rm el}$ does not appear explicitly in Eqs.~\eqref{eq:BoltzPhi_chi}--\eqref{eq:BoltzPhi}.

Keeping the \emph{number density} near equilibrium, $n_\phi\simeq n_{\phi,{\rm eq}}$, additionally requires \emph{chemical} processes, such as decay/inverse decay of $\phi$ and annihilation into SM particles, to be faster than the Hubble rate $H$:
\begin{equation}
\Gamma_{\rm chem}(T)\;\equiv\;\langle\Gamma_\phi\rangle(T)\;+\;n_\phi^{\rm eq}(T)\,\langle \sigma v_{\rm{rel}} \rangle_{\phi\phi\to{\rm SM\,SM}}\ \gtrsim\ H(T)\,,
\end{equation}
(where for a complex $\phi$ one replaces $n_\phi^{\rm eq\,2}\!\to n_\phi^{\rm eq}\,n_{\phi^\dagger}^{\rm eq}$)\footnote{If $\phi$ is self--conjugate, a symmetry factor $1/2$ can be included either in the definition of $\langle \sigma v_{\rm{rel}} \rangle_{\phi\phi\to{\rm SM}}$ or as a prefactor of the corresponding term in Eq.~\eqref{eq:BoltzPhi}. If $\phi$ is complex and you track $n_\phi$ and $n_{\phi^\dagger}$ separately, replace $n_\phi^2 \to n_\phi n_{\phi^\dagger}$ and likewise for $n_{\phi,{\rm eq}}^2$.}
Thus, elastic scattering alone is not sufficient to keep the mediator $\phi$ in thermal equilibrium with the bath; decays/inverse decays and annihilation/inverse annihilation must also be included when assessing whether $\phi$ remains in equilibrium with the SM bath.

If $\phi$ remains in thermal contact with the SM bath, i.e.\ with both chemical and kinetic equilibrium, the coupled Boltzmann system reduces to the single Lee--Weinberg equation (Eq.~\ref{eq:leeweinberg}) for $n_\chi$, with
\begin{equation}
\langle \sigma v_{\rm{rel}} \rangle \;=\; \langle \sigma v_{\rm{rel}} \rangle_{\chi\bar\chi\to \phi\phi}
\;+\; \langle \sigma v_{\rm{rel}} \rangle_{\chi\bar\chi\to {\rm SM}} \, ,
\end{equation}
with the same temperature for the SM and hidden sector.
Subsequent $\phi$ decays repopulate SM states while leaving the relic density $\Omega_{\rm DM} h^2$ determined by $\chi\bar\chi\to \phi\phi$ (provided late decays do not significantly inject entropy).

In our benchmarks the $\phi$ decay rate typically exceeds $\phi\phi\to{\rm SM}$; this holds in all three models.
Therefore, the $\phi$ decay rate is the dominant process that keeps this particle in the chemical equilibrium (see Fig.~\ref{fig:modelIrate}).
The thermally averaged decay rate scales with temperature as:
\begin{equation}
\langle \Gamma_\phi\rangle(T)\simeq
\begin{cases}
\Gamma_\phi\,\dfrac{m_\phi}{2T}\;\propto T^{-1}\,, & T\gg m_\phi\,,\\[8pt]
\Gamma_\phi \quad\text{(constant)}\,, & T\ll m_\phi\,.
\end{cases}
\end{equation}
Therefore, when the mediator becomes non -relativistic the thermally average decay rate becomes constant with temperature.
Since $H(T)\propto T^2$ during radiation domination,
\begin{equation}
\frac{\langle \Gamma_\phi\rangle}{H}\propto
\begin{cases}
T^{-3}\,, & T\gg m_\phi\,,\\[4pt]
T^{-2}\,, & T\ll m_\phi\,,
\end{cases}
\end{equation}
so $\langle \Gamma_\phi\rangle/H$ \emph{grows} as the Universe cools. 
Therefore, if
\begin{equation}
\frac{\langle\Gamma_\phi\rangle}{H}\Big|_{T\sim m_\phi}\gtrsim 1,
\end{equation}
then $\langle\Gamma_\phi\rangle/H\gg 1$ also at $T\sim m_\chi/x_f$, even though $n_\phi$ is Boltzmann suppressed. 
Therefore, a sufficient (model--dependent) condition to control whether $\phi$ particles are in equilibrium is that $\Gamma_{\rm chem}(T)$
exceed the Hubble rate at $T\sim \max(m_\phi,\,m_\chi/x_f)$, with $x_f\simeq 20$--$30$ for weak-scale WIMPs.

If kinetic equilibrium is not maintained ($\Gamma_{\rm el} (T_f) \lesssim H(T_f)$) it is still fine to solve a single Boltzmann equation for $\chi$, but the dark sector follows its own temperature $T'$. 
If, instead, also chemical equilibrium is not kept during freeze--out, the coupled system in Eqs.~\eqref{eq:BoltzPhi_chi}--\eqref{eq:BoltzPhi} must be solved. In the next sections we will report the typical values of the couplings for which we can assume the particle $\phi$ is in equilibrium with the bath.

The processes that produce DM from the mediator, such as
$\phi \to \chi\bar{\chi}$ and $\phi\phi \to \chi\bar{\chi}$, are not relevant in the secluded regime ($m_\chi > m_\phi$) to decide whether the particle $\phi$ is in thermal equilibrium with the bath. In fact, the decay
$\phi \to \chi\bar{\chi}$ is kinematically forbidden for $m_\phi < 2m_\chi$, while the annihilation $\phi\phi \to \chi\bar{\chi}$ is exponentially Boltzmann suppressed at $T \sim m_\chi$ (it requires two energetic $\phi$’s from the thermal tail to overcome the threshold $2m_\chi$). Moreover, $\phi$--to--$\chi$ conversion processes only equilibrate the \emph{dark sector} ($\chi$--$\phi$) and contribute to bath contact \emph{only} if $\chi$ itself is in equilibrium with the SM at the relevant temperature (e.g.\ via $\chi\bar{\chi}\leftrightarrow f\bar{f}$). This condition is not satisfied in the secluded regime, where $\chi$ does not couple directly to SM fermions.

\subsubsection{Model I ($U(1)_X$ with Dirac DM, dark photon and kinetic mixing)}

For $m_\chi>m_{A'}$ the dominant secluded process is $\chi\bar\chi\to A'A'$ (via $t/u$ channels) (see Eq.~\ref{eq:mod1sec}), while $A'$ maintains equilibrium through kinetic mixing, parametrized by $\epsilon$, with the hypercharge/EM current. 

\begin{figure}
\includegraphics[width=0.99\linewidth]{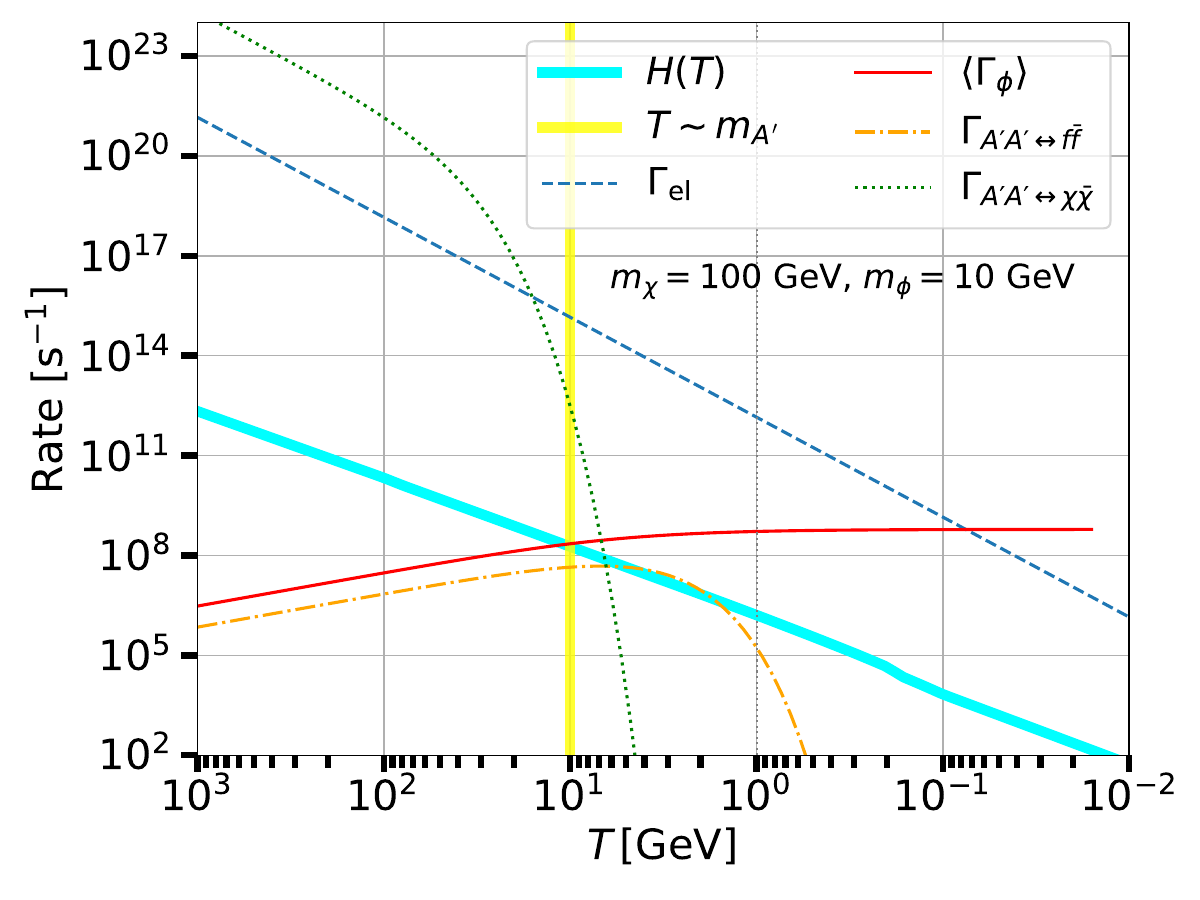}
\caption{Rates for the processes that control chemical and kinetic equilibrium. We show the rate for $A'$ decay $\langle \Gamma_{A'} \rangle$ (red solid), annihilation into fermions $\Gamma_{A'A'\leftrightarrow f\bar{f}}$ (orange dot--dashed) and elastic scattering $\Gamma_{\rm el}$ (blue dashed). We also display the rate for $A'A' \to \chi \bar{\chi}$ (green dotted), the Hubble rate (cyan band) and the temperature equivalent to the dark photon mass (yellow band).}
\label{fig:modelIrate}
\end{figure}

We show in Fig.~\ref{fig:modelIrate} the rates $\Gamma(T)$ for the processes that control the \emph{chemical} and \emph{kinetic} equilibrium of the dark photon $A'$ with the SM bath, together with the rate that couples $A'$ to the DM particle $\chi$. We fix $m_\chi=100~\mathrm{GeV}$, $m_{A'}=10~\mathrm{GeV}$, $\epsilon=5\times 10^{-8}$ and $g_X=0.034$, chosen to reproduce the observed relic abundance of $\chi$.

The annihilation rate involving DM, $A'A'\!\to\!\chi\bar\chi$, is \emph{not} the dominant contribution for $T\lesssim 10~\mathrm{GeV}$: in the secluded case ($m_{A'}\ll m_\chi$) this channel is strongly Boltzmann suppressed at $T\ll m_\chi$ and only receives support from the high--energy tail of the $A'$ distribution. Its contribution therefore falls rapidly below $T\sim m_\chi$ and does not control the chemical contact of $A'$ with the SM at the temperatures shown.

Below $10~\mathrm{GeV}$ and down to about $0.1~\mathrm{GeV}$, the elastic scattering becomes the dominant process, with a rate much larger than the Hubble one, $H(T)$. This controls the kinetic equilibrium of the dark photon with the SM bath which is maintained across the relevant temperatures for DM freeze-out.

As for the processes that give chemical equilibrium, the decay and inverse decay provide the largest rate, which is larger than $H(T)$ for $T\lesssim m_{A'}$. 
At high temperatures ($T\gg m_{A'}$) the thermally averaged decay rate
$\langle \Gamma_{A'}\rangle=\Gamma_{A'}\,K_1(m_{A'}/T)/K_2(m_{A'}/T)$ is suppressed as
$\langle \Gamma_{A'}\rangle \sim \Gamma_{A'}\,(m_{A'}/2T)$, while the Hubble rate increases with temperature as $H\propto T^2$. As the universe cools to $T\lesssim m_{A'}$, one has $K_1/K_2\to 1$, so $\langle\Gamma_{A'}\rangle\to \Gamma_{A'}$ (constant) and $\langle\Gamma_{A'}\rangle/H$ grows, crossing unity near $T\sim m_{A'}$. This indicates that $A'$ is not in chemical equilibrium at very high $T$, but it \emph{is} in chemical equilibrium with the SM for $T\lesssim m_{A'}$, which is the most relevant part for the DM freeze--out that occurs around a few GeV.

In summary, although $A'$ is not in chemical equilibrium at very high temperatures, it \emph{recouples} near $T\sim m_{A'}$ and remains in both chemical and kinetic equilibrium with the SM at the epoch of DM freeze--out ($T_f\sim 4~\mathrm{GeV}$) for the benchmark shown. Consequently, the relic density can be computed by solving the single Lee--Weinberg equation (Eq.~\ref{eq:leeweinberg}) for $n_\chi$ with $T'=T$, while $A'$ subsequently decays to SM particles without affecting $\Omega_{\rm DM}h^2$ (absent significant late entropy injection).

For values of $\epsilon$ larger than $5\times 10^{-8}$ the decay rate crosses the Hubble rate at temperatures larger than $m_{A'}$ therefore $A'$ is already in chemical equilibrium above  $T\sim m_{A'}$ and certainly throughout freeze-out.

We can now numerically find the minimum value of $\epsilon$ for which $A'$ is in chemical and kinetic equilibrium.
As shown in Fig.~\ref{fig:modelIrate}, the elastic scattering rate is larger than the processes that provide chemical equilibrium for $T$ between $m_{A'}$ and $T_f$, therefore if we choose $\epsilon$ to keep chemical equilibrium we automatically have kinetic equilibrium.
For $m_{A'}>2m_e$ the most important $A'$--number--changing process is the decay and inverse decay, whose width and thermal average are
\begin{eqnarray}
&& \Gamma_{A'} = \epsilon^2 \alpha\, m_{A'}\sum_{f}\frac{N_c^f Q_f^2}{3}\left(1+2\frac{m_f^2}{m_{A'}^2}\right)\sqrt{1-4\frac{m_f^2}{m_{A'}^2}}\,,
\\ 
&& \langle\Gamma_{A'}\rangle =\Gamma_{A'}\,\frac{K_1(m_{A'}/T)}{K_2(m_{A'}/T)}\,. \nonumber
\end{eqnarray}
Requiring $\langle\Gamma_{A'}\rangle \gtrsim H$ at $T\simeq m_{A'}$ gives
\begin{eqnarray}
&&\epsilon \gtrsim
\left[
\frac{1.66\,\sqrt{g_\ast}}{\tfrac{\alpha}{3}}\;
\frac{m_{A'}}{M_{\rm Pl}}\;
\frac{1}{(K_1/K_2)|_{x_{A'}=1}}
\right]^{\!1/2}  \approx \\ 
&& \approx 5\times 10^{-8}\;
\left(\frac{g_\ast}{80}\right)^{\!1/4}
\left(\frac{m_{A'}}{10~\mathrm{GeV}}\right)^{\!1/2}\!. 
\label{eq:eps_equil}
\end{eqnarray}
Therefore, for $m_{A'}=10~\mathrm{GeV}$ a kinetic mixing of order $\epsilon \sim 5 \times 10^{-8}$ is sufficient to keep the dark photon $A'$ in chemical equilibrium with the bath as shown in Fig.~\ref{fig:modelIrate}. This value is much smaller than the upper limits coming from collider experiments \cite{Bauer:2018onh,Koechler:2025ryv}. Moreover, as we will see in Sec.~\ref{sec:results}, such small values of the kinetic mixing suppress significantly the nuclear cross sections that are proportional to $\epsilon^2$ (see Eq.~\ref{sec:mod1DD}).

\subsubsection{Model II ({\tt DMSimp} with scalar DM $S$ and scalar mediator $K$)}
For $m_S>m_K$, the secluded process $S S \to K K$ controls freeze--out, while $K$ stays in contact with the SM via its Yukawa--like coupling $\propto \beta\,m_f/v_h$. 
The $K$ decay width into fermions is
\begin{eqnarray}
\Gamma(K\!\to f\bar f)= \sum_f N_c^f\;\frac{\beta^2\,m_f^2\,m_K}{8\pi\,v_h^2}\;
\left(1-\frac{4m_f^2}{m_K^2}\right)^{3/2},
\label{eq:GammaY}
\end{eqnarray}
where the sum is performed on the decay channels kinematically open.
Demanding $\langle\Gamma_K\rangle \gtrsim H$ at $T\simeq m_K$ yields the qualitative bound
\begin{eqnarray}
\beta \gtrsim
\left[
\frac{1.66\,\sqrt{g_\ast}}{N_c^f}
\frac{8\pi\,v_h^2}{m_f^2}
\frac{1}{M_{\rm Pl}\,m_K}
\frac{1}{K_1/K_2}\Big|_{x_{K}=1}
\right]^{\!1/2} \simeq \\
\simeq
2\times 10^{-6}\;
\left(\frac{g_\ast}{80}\right)^{\!1/4}
\frac{1}{\sqrt{N_c^f}}
\left(\frac{1~\mathrm{GeV}}{m_f}\right)
\sqrt{\frac{1~\mathrm{GeV}}{m_K}}\,, \nonumber
\label{eq:betaS_num}
\end{eqnarray}
where $N_c^f=1$ for leptons and $N_c^f=3$ for quarks, and $m_f$ is the mass of the \emph{heaviest} kinematically open fermion ($m_f<m_Y/2$), which maximizes $\Gamma(Y\to f\bar f)$.
%Elastic scattering $Y f\!\leftrightarrow\! Y f$ via the same coupling further helps maintain equilibrium; its rate can be estimated as $\Gamma_{\rm scatt}\sim n_f\langle \sigma v_{\rm{rel}} \rangle\propto \beta^2 y_f^2 T$ and added to $\langle\Gamma_Y\rangle$ in the criterion $\Gamma_\phi^{\rm bath}\gtrsim H$.

Elastic scattering $K f\!\leftrightarrow\! K f$ via the Yukawa $y_f=\beta\,m_f/v_h$ helps maintain \emph{kinetic} equilibrium.
Its momentum–transfer rate scales as $\Gamma_{\rm el}^{(Kf)} \sim n_f^{\rm eq}\,\langle\sigma_T v\rangle \propto y_f^{4}T
= \beta^{4}\Big(\tfrac{m_f}{v_h}\Big)^{4} T,$
up to order-one angular/phase–space factors.  This term should be added to the bath–coupling rate when testing $\Gamma_{\rm bath}\gtrsim H$, but it does not change number densities because for $\beta\ll 1$ it is highly suppressed with respect to the decay rate in Eq.~\ref{eq:GammaY}.

\subsubsection{Model III ($U(1)_X$ with Dirac DM with a Higgs--mixed singlet that generates $m_\chi$)}
For $m_\chi>m_{H_p}$, $\chi\bar\chi\to H_p H_p$ is the secluded channel that dominates the relic density processes. 
The mediator $H_p$ couples to the SM through Higgs mixing $\sin\alpha$ and its total width is well approximated by
\begin{equation}
\Gamma_{H_p}\simeq \sin^2\alpha\;\Gamma_h^{\rm SM}(m_h\!\to\! m_{H_p}),
\end{equation}
where $\Gamma_h^{\rm SM}$ is the SM Higgs decay width.
The equilibrium condition at $T\simeq m_{H_p}$ is then
\begin{eqnarray}
&\sin&\alpha \gtrsim
\sqrt{\frac{1.66\,\sqrt{g_\ast}\,m_{H_p}}{M_{\rm Pl}\,\Gamma_h^{\rm SM}(m_{H_p}) \frac{1}{K_1/K_2}\Big|_{x_{H_p}=1}}} \sim \\
 &\sim& 1.8\times 10^{-6}\;
\left(\frac{g_\ast}{80}\right)^{\!1/4} \frac{m_{H_p}}{10 \,\mathrm{GeV}}
\left(\frac{\Gamma_h^{\rm SM}(m_{H_p})}{\mathrm{MeV}}\right)^{-1/2} \nonumber
\label{eq:mix_equil}
\end{eqnarray}
Therefore, for a 10 GeV mediator the lower limit for $\sin\alpha$ typically corresponds to $\sin\alpha\gtrsim 2\times 10^{-6}$, with hadronic uncertainties near thresholds.

In this model the $H_p$--fermion Yukawa coupling is $y_f=(m_f/v)\sin\alpha$, so the
momentum–transfer (elastic) rate scales as
\begin{equation}
\Gamma_{\rm el}(T)\sim\sum_f n_f^{\rm eq}(T)\langle\sigma_T v\rangle
\propto \sin^{4}\alpha\sum_f\left(\frac{m_f}{v}\right)^{4}T,
\end{equation}
(dominated by the heaviest relativistic fermions, e.g.\ $b,\tau$ near $T\!\sim\!{\rm few}$–$10$ GeV).
By contrast, the chemical rate from decays scales only as
\(\langle\Gamma_{H_p}\rangle\simeq \sin^{2}\!\alpha\,\Gamma_h^{\rm SM}(m_{H_p})\,K_1/K_2\),
so for small mixing \(\Gamma_{\rm dec}\gg\Gamma_{\rm el}\); elastic becomes comparable only at much larger \(\sin\alpha\) because of the extra \(\sin^{2}\!\alpha\) suppression.

%the BBN safety requirement $\tau_\phi\!\lesssim\!1\,\mathrm{s}$ gives a numerically similar or slightly weaker lower bound.

Therefore, in all three models considered, the dark-sector mediator or scalar remains in chemical equilibrium for coupling values to SM particles that are much smaller than unity. 
These same couplings also determine the nuclear cross sections relevant for direct detection. 
Consequently, in the secluded regime, one can significantly reduce the size of these couplings while still ensuring that the dark-sector mediator or scalar stays in equilibrium with the thermal bath. 
This justifies solving the Lee--Weinberg equation, Eq.~\ref{eq:leeweinberg}, by tracking only the evolution of the DM abundance.

\begin{figure}
\includegraphics[width=0.99\linewidth]{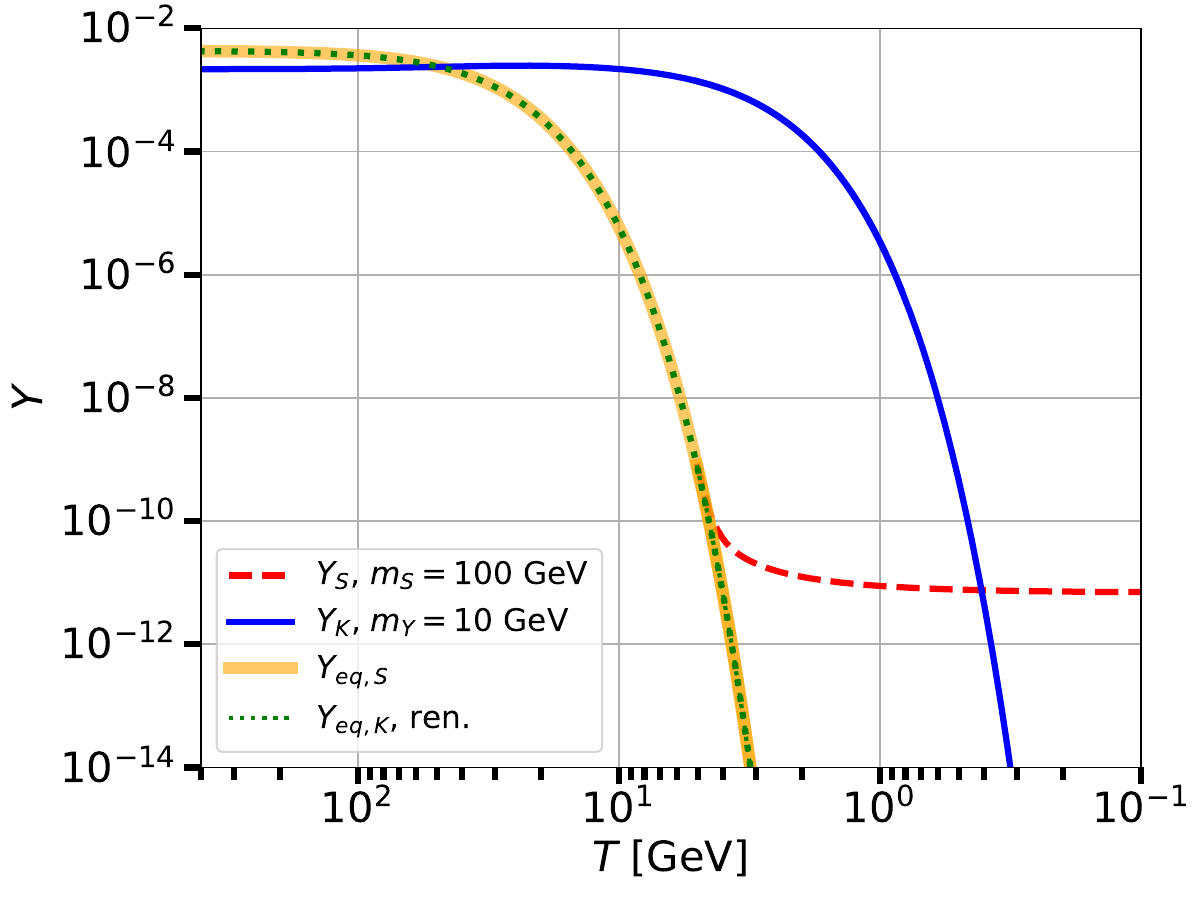}
\caption{This plot shows the comoving number density $Y$ for the DM particle $S$ and mediator $K$ for Model II when fixing $m_K=10$ GeV and $m_S=100$ GeV and $g_K=0.8$ and $\beta=0.05$. We are thus in the secluded regime and the dominant annihilation channel for these parameters is $SS \to KK$. We show also the equilibrium $Y$ for the two particles and the rescaled equilibrium number density for the particle $K$ ($Y_{\rm{eq},K}$) but rescaled in order to be directly comparable with the DM one $Y_{\rm{eq},S}$.}
\label{fig:Y}
\end{figure}

We further demonstrate our assumption by comparing the value of $\Omega_{\rm{DM}}h^2$ and the shape of the comoving number density obtained by solving the coupled Boltzmann equation in Eq.~\ref{eq:BoltzPhi} and \ref{eq:BoltzPhi_chi} for DM and mediator  with the one obtained when using only the single Lee-Weinberg equation for the DM particle. We employ the {\tt micrOMEGAs} code for this task \cite{Belanger:2006is,Belanger:2013oya,Barducci:2016pcb,Belanger:2018ccd}. We perform this exercise for the Model II where we have the scalar mediator $K$ and scalar DM $S$.
We fix $m_K=10$ GeV, $m_S=100$ GeV, $g_K=0.8$ and $\beta=0.05$. This choice permits to reach the observed relic density for $S$ when solving the Lee-Weinberg Eq.~\ref{eq:leeweinberg}. We obtain a compatible value below the $\%$ level also when following the $S$ and $K$ densities by means of Eq.~\ref{eq:BoltzPhi} and \ref{eq:BoltzPhi_chi}. This is due to the fact that with $\beta=0.05$ the mediator $K$ is in thermal and chemical equilibrium with the bath and so the two approaches must provide the same result.

We show in Fig.~\ref{fig:Y} the comoving number density $Y$ as a function of the temperature for the DM particle $S$ ($Y_S$) and mediator $K$ ($Y_K$). We show also the equilibrium quantity $Y_{\rm{eq}}$ for the two particles. 
We note that for temperatures larger than the particle masses $Y$ is basically flat. This is due to the fact that for relativistic particles $n \propto T^3$ as well as the entropy density $s$ so $Y_{\rm{eq}}=n_{\rm{eq}}/s$ does not change with $T$.
Instead, for $T<100$ GeV ($T<10$ GeV) the DM (mediator) particle equilibrium comoving number density decreases exponentially because it becomes non-relativistic and $n_{\rm{eq}}\propto T^3 \exp{(-m/T)}$ so $Y_{\rm{eq}}\propto T^{3/2} \exp{(-m/T)}$.

$Y_{\rm{eq}}$ is given for a particle with internal degrees of freedom $g$ by \cite{KolbTurner,GONDOLO1991145}
\begin{equation}
Y_{\text{eq}}(T) \;=\; \frac{n^{\text{eq}}(T)}{s(T)} 
= \frac{45}{4\pi^4} \, \frac{g}{g_\ast(T)} \, x^2 K_2(m/T),
\end{equation}
where \(K_2(x)\) is the modified Bessel function of the second kind.
Two particles like $S$ and $K$ would naturally have different $Y$ if they have different masses and internal degrees of freedom. In order to overlay one on the other in Fig.~\ref{fig:Y} and verify that for the choice of the model parameters the mediator $K$ follows the same $Y_{\rm{eq}}$ as the DM particle, we can consider the ratio
\begin{equation}
\label{eq:ratioY}
\frac{Y_{\text{eq},S}(x)}{Y_{\text{eq},K}(x)}
= \frac{g_S}{g_K} r^2 \, \frac{K_2(r x)}{K_2(x)}.
\end{equation}
where $r \equiv \frac{m_S}{m_K}$.
%In the non-relativistic regime (\(x \gg 1\)), one may use the asymptotic expansion
%\begin{equation}
%K_2(x) \;\approx\; \sqrt{\frac{\pi}{2x}} \, e^{-x}.
%\end{equation}
In the non-relativistic regime (\(x \gg 1\)) the ratio simplifies to
\begin{equation}
\frac{Y_{\text{eq},S}(x)}{Y_{\text{eq},K}(x)}
\;\approx\; r^{3/2} \, e^{-x(r-1)}.
\end{equation}
%For the case shown in the figure with (\(r=10\)), we obtain a ratio of $Y_{\text{eq},S}(x)/Y_{\text{eq},Y}(x) \approx  31.6 \, e^{-9x}$. 
When we apply the rescaling as in Eq.~\ref{eq:ratioY} we obtain that the equilibrium number density for the DM particle and the rescaled one for the mediator are approximately the same. This justifies our assumption of solving the Lee-Weinberg equation instead of the coupled Boltzmann equation for $S$ and $K$.

\subsection{Big Bang Nucleosynthesis constraints}
\label{sec:BBN}

To avoid spoiling BBN, any metastable dark–sector mediator present in the early Universe must decay well \emph{before} the BBN epoch; a conservative requirement applicable to all three models is
\[
\tau_\phi \lesssim\; 1~\text{s} \,\, (T\gtrsim 1~\text{MeV})
\Longleftrightarrow
\Gamma_\phi \gtrsim 1~\text{s}^{-1} \simeq 6.6\times 10^{-25}\ \text{GeV},
\]
with a more stringent criterion \(\tau_\phi \lesssim 0.1~\text{s}\) often adopted for hadronically dominated decays or scenarios with significant late electromagnetic injection.
Below we translate this into bounds on the portal couplings of our three models.

\subsubsection{Model I ($U(1)_X$ Dirac \(\chi\) with dark photon \(A'\) and kinetic mixing)}

Demanding \(\Gamma_{A'}\ge \Gamma_{\rm BBN}\) and far above threshold $m_{A'}>2m_f$ gives
\begin{eqnarray}
\epsilon \gtrsim 
5.2\times 10^{-11}\,
\sqrt{\frac{1~\mathrm{GeV}}{m_{A'}}}\;
\left[\frac{1}{\sum\limits_{f\;{\rm open}} N_c^f Q_f^2\,}\right]^{\!1/2},
\end{eqnarray}
where the sum is performed on all the kinematically open decays into fermions.
%A conservative (channel–by–channel) lower bound follows by keeping only the heaviest \emph{open} fermion(s) in the sum. 
%For \(m_{A'}\gg 2m_e\) and electrons only,
%\[
%\epsilon \;\gtrsim\; 1.6\times 10^{-11}\;\sqrt{\frac{1~\text{GeV}}{m_{A'}}}\,,
%\]
%which is further relaxed when \(\mu^+\mu^-\) and hadronic modes open (they increase \(\sum_f N_c^f Q_f^2\Phi_f\)).

\subsubsection{Model II ({\tt DMSimp} scalar DM $S$ with scalar mediator $K$)}

In this model forcing \(\Gamma_K\ge \Gamma_{\rm BBN}\) implies
\begin{equation}
\beta \gtrsim
\frac{1.0\times 10^{-9}}{\sqrt{N_c^f}}\;
\left(\frac{1~\text{GeV}}{m_f}\right)\,
\sqrt{\frac{1~\text{GeV}}{m_K}}\;
\left(1-\frac{4m_f^2}{m_K^2}\right)^{-\!3/4}
\end{equation} 
We should use the \emph{heaviest kinematically open} fermion \(f\) to obtain the weakest (most permissive) bound.

\subsubsection{Model III ($U(1)_X$ Dirac \(\chi\) with Higgs–mixed singlet \(H_p\))}

Here \(\Gamma_{H_p}=\sin^2\alpha\;\Gamma_h^{\rm SM}(m_{H_p})\), where \(\Gamma_h^{\rm SM}(m)\) is the SM Higgs width evaluated at mass \(m\). The BBN condition gives
\begin{equation}
\sin\alpha \gtrsim \sqrt{\frac{\Gamma_{\rm BBN}}{\Gamma_h^{\rm SM}(m_{H_p})}} \sim 8.1 \times 10^{-11} \sqrt{\frac{\text{MeV}}{\Gamma_h^{\rm SM}(m_{H_p})}}
\end{equation}
which weakens rapidly above hadronic and electroweak thresholds where \(\Gamma_h^{\rm SM}(m_{H_p})\) grows.

%\vspace{0.5em}
%\emph{Rescaling for a different BBN time cut.} If you adopt \(\tau_\phi^{\rm max}\neq 1~\text{s}\), replace \(\Gamma_{\rm BBN}\) by \(6.58\times 10^{-25}\,\text{GeV}\times (1~\text{s}/\tau_\phi^{\rm max})\). Since \(\Gamma\propto \epsilon^2,\beta_S^2,\sin^2\alpha\), the bounds on \(\epsilon,\beta_S,\sin\alpha\) scale as \(\propto (\tau_\phi^{\rm max})^{-1/2}\).

For all the three models, the lower limits obtained for the portal coupling constants to make the mediator or scalar $H_p$ to decay before BBN are much smaller than then one obtained in the previous section to keep them in chemical equilibrium with the SM bath.

\section{Results}\label{sec:results}

In this section we report the results for the three models we consider.
We impose the following constraints:
\begin{itemize}
\item \textbf{Cosmology:} the relic abundance should match $\Omega_{\rm DM} h^2 \simeq 0.120$ as reported by Ref.~\cite{Planck:2018vyg}.
\item \textbf{Direct detection:} nuclear cross sections should be consistent with the upper limits on DM–nucleus scattering from LZ~\cite{LZ:2024zvo}.
\item \textbf{Indirect detection:} annihilation cross sections must comply with the upper limits on $\langle \sigma v_{\rm{rel}} \rangle$ from the combined $\gamma$-ray analysis of dSphs~\cite{McDaniel:2023bju}.
\end{itemize}

For each Model we generate the associated {\sc UFO}~\cite{Degrande:2011ua} and {\sc CalcHEP}~\cite{Belyaev:2012qa} files with {\sc FeynRules}~\cite{Alloul:2013bka}, and pass them to \maddm~\cite{Backovic:2013dpa,Ambrogi:2018jqj,Arina:2021gfn} to compute the prompt $\gamma$-ray yields used to derive coupling limits from the combined dSph analysis in \emph{Fermi}-LAT data~\cite{McDaniel:2023bju}. We use the source spectra from \texttt{Cosmixs} \cite{Arina:2023eic,DiMauro:2024kml}.
We use \micromegas~\cite{Belanger:2006is,Belanger:2013oya,Belanger:2018ccd,Alguero:2023zol} to evaluate the relic density and direct-detection signals.
We use \micromegas\ {\tt 6.0.4} and \maddm\ {\tt 3.2}.

\subsection{Model I ($U(1)_X$ with kinetic mixing)}

\subsubsection{WIMP regime}

\begin{figure*}
  \centering
  \includegraphics[width=0.49\linewidth]{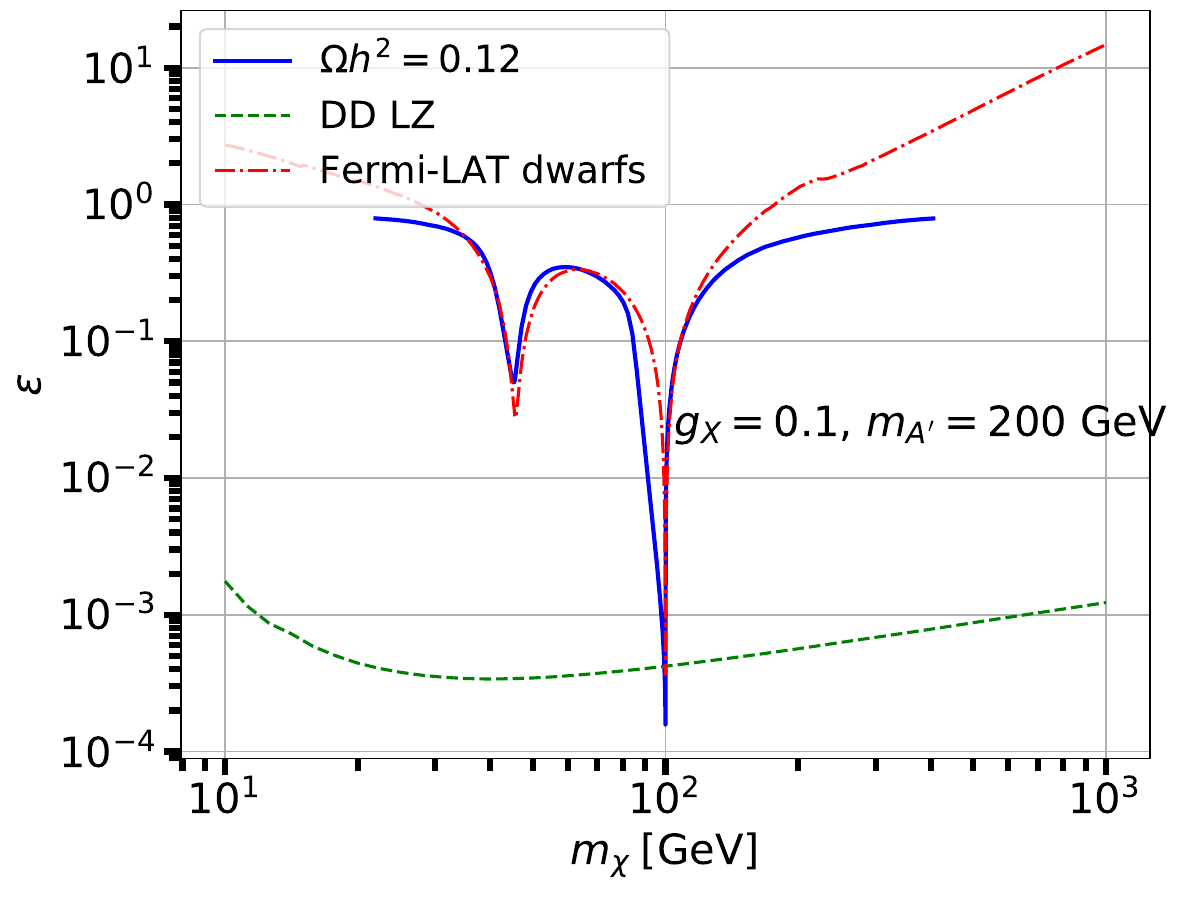}
  \includegraphics[width=0.49\linewidth]{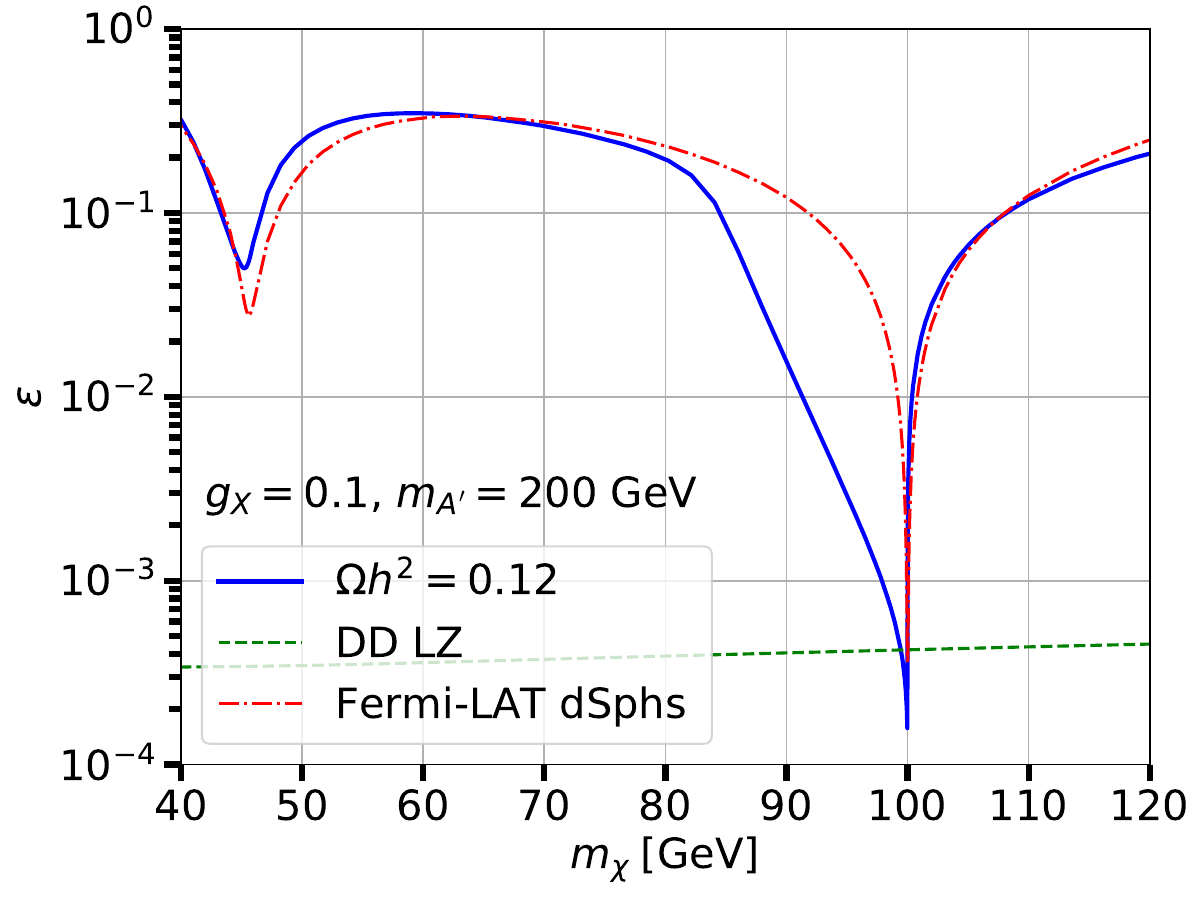}
  \includegraphics[width=0.49\linewidth]{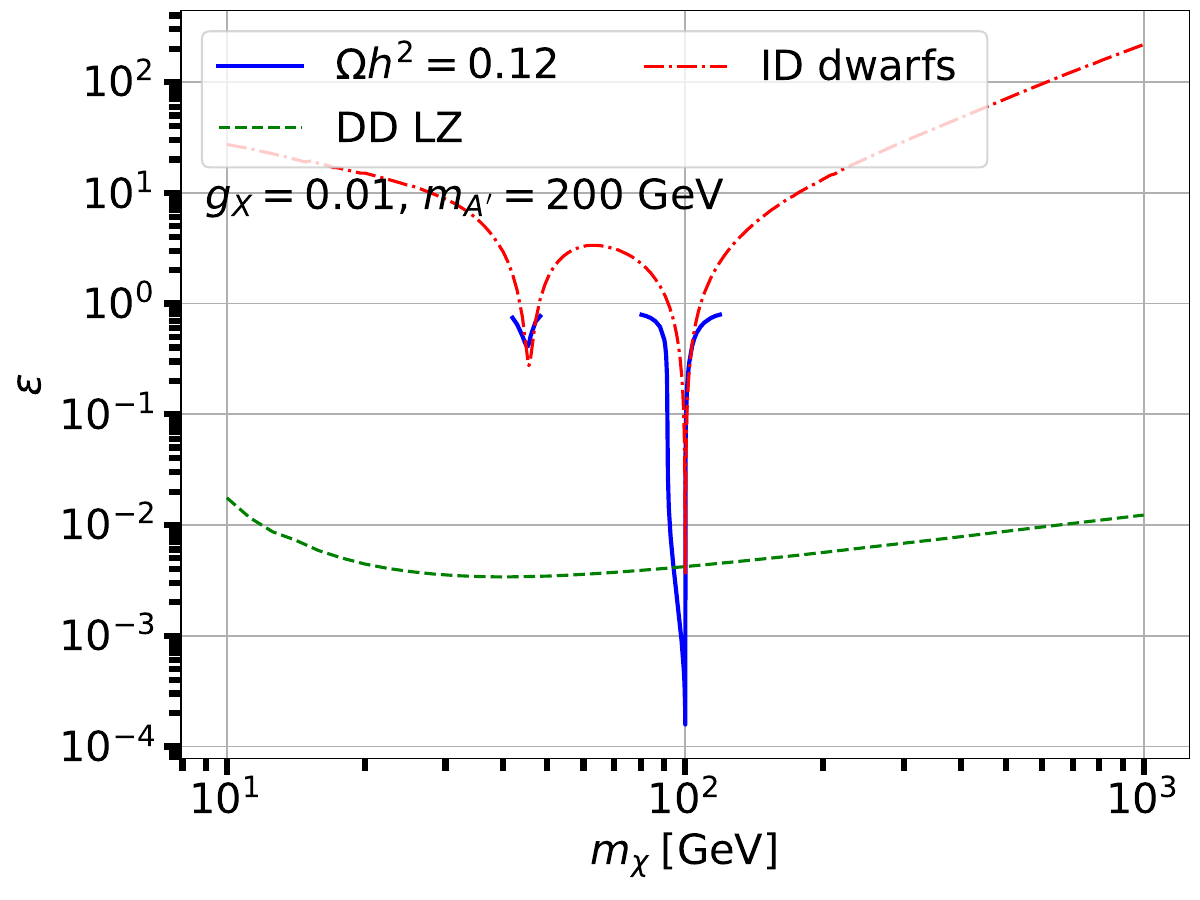}
  \includegraphics[width=0.49\linewidth]{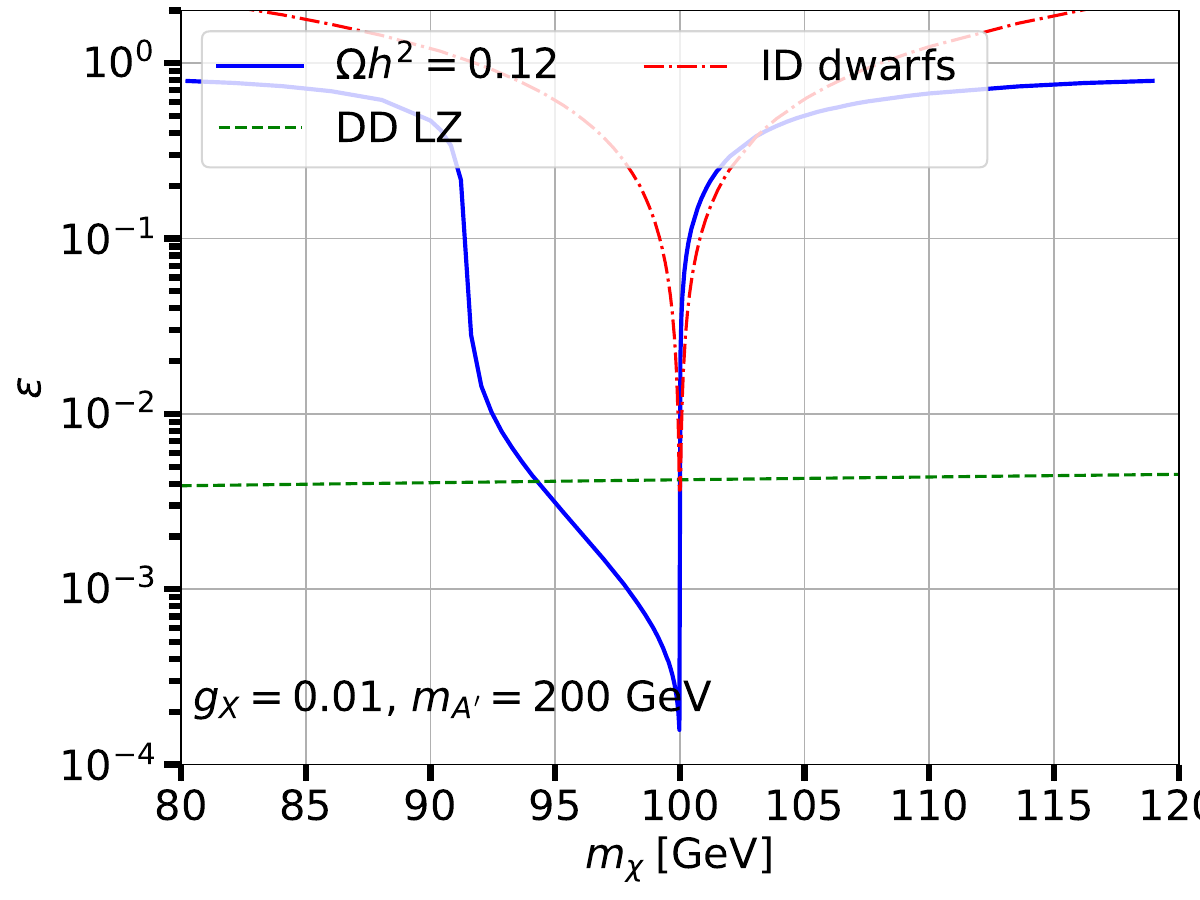}
  \includegraphics[width=0.49\linewidth]{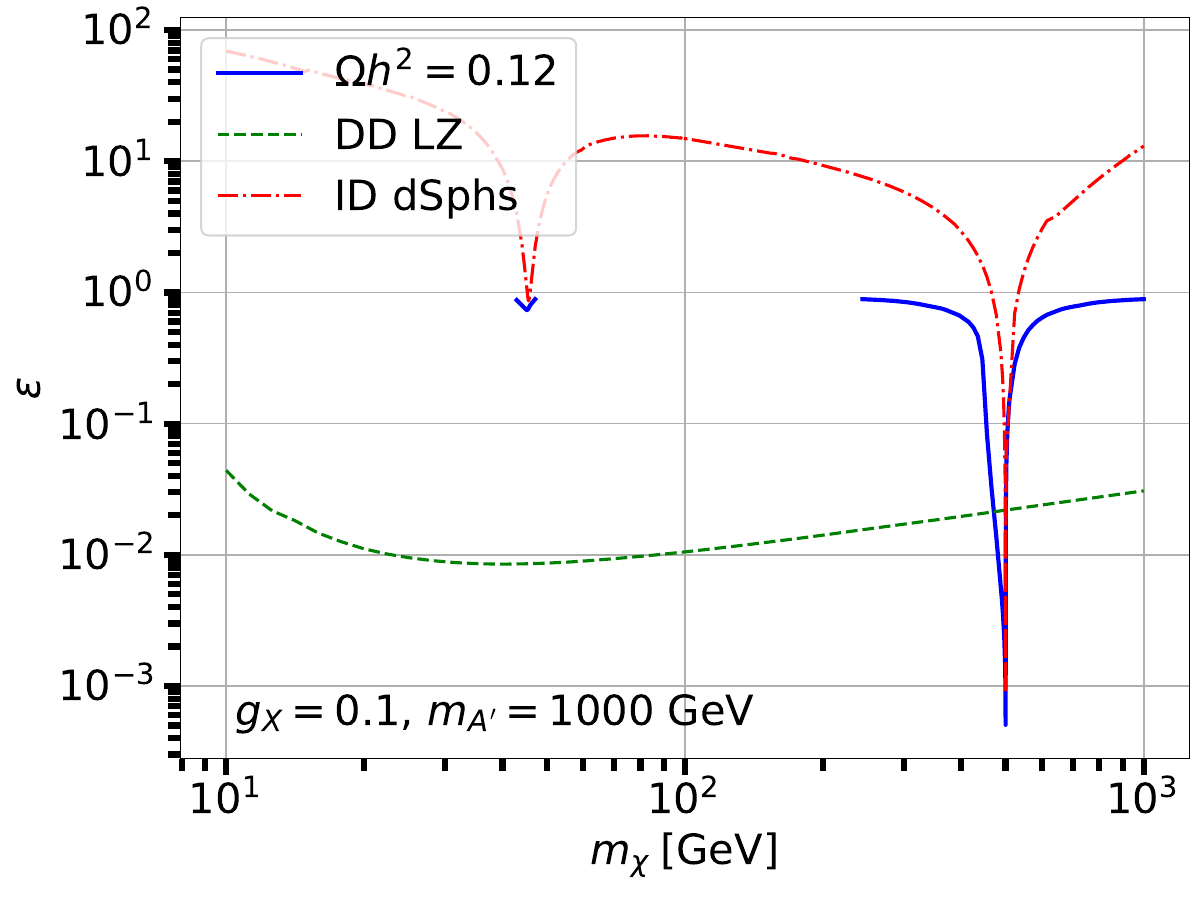}
  \includegraphics[width=0.49\linewidth]{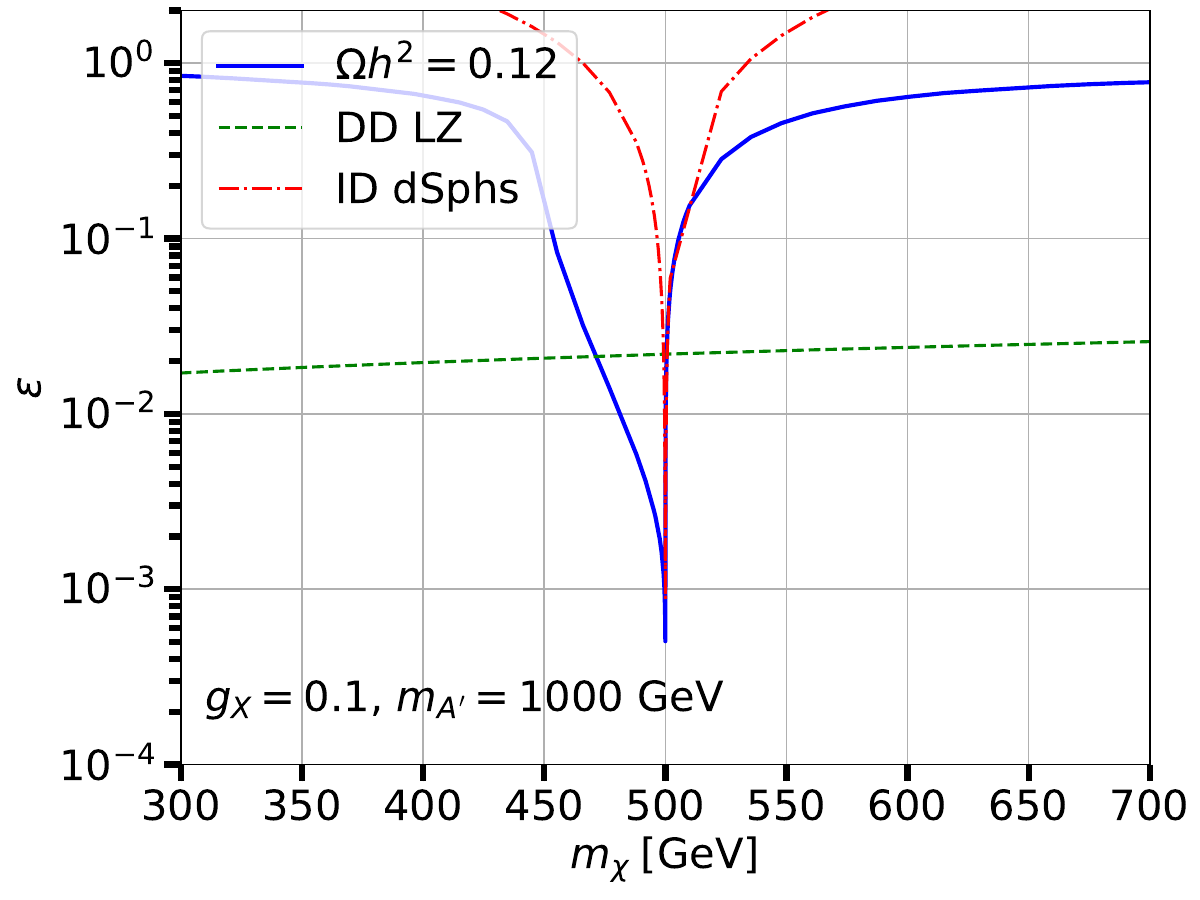}
  \caption{This figure shows the $(m_\chi,\epsilon)$ values that reproduce the observed relic abundance (blue solid). Also shown are the upper limits on $\epsilon$ inferred from the LZ spin–independent WIMP–nucleon bound (green dashed) and from \emph{Fermi}–LAT dwarf–spheroidal $\gamma$–ray constraints (red dot–dashed).}
\label{fig:U1XWimp}
\end{figure*}

We show in Fig.~\ref{fig:U1XWimp} the results for the \emph{WIMP} regime, in which the relic density is dominated by annihilation into SM fermions via kinetic mixing. 
We consider $m_{A'}=200~\mathrm{GeV}$ and $1~\mathrm{TeV}$, scan $m_\chi\in[10~\mathrm{GeV},\,1~\mathrm{TeV}]$, and fix representative values of the dark gauge coupling $g_X=\{0.1,\,0.01\}$.\footnote{For $g_X=10^{-3}$ no point in the scanned range reproduces $\Omega_{\rm DM}h^2\simeq0.12$, so we do not show it in Fig.~\ref{fig:U1XWimp}.}
For each point we determine the portal coupling $\epsilon$ that reproduces the observed relic density within its experimental uncertainty~\cite{Planck:2018vyg}, and confront it with the LZ spin–independent limit~\cite{LZ:2024zvo} and with the \emph{Fermi}–LAT dSphs bound~\cite{McDaniel:2023bju}.

The relic–density curves \emph{exhibit} two characteristic dips at $m_\chi\simeq m_Z/2$ and $m_\chi\simeq m_{A'}/2$. 
These arise from $s$–channel resonances that enhance $\chi\bar\chi\!\to f\bar f$ near half the mediator masses (SM $Z$ and dark photon $A'$), sharply increasing $\langle \sigma v_{\rm{rel}} \rangle$ and thus reducing the $\epsilon$ required to achieve $\Omega_{\rm DM}h^2\simeq0.12$.
Away from both resonances one typically needs $\epsilon=\mathcal{O}(0.1{-}1)$ to match the relic density, as expected for GeV–TeV WIMPs and cross sections of the order of the electroweak strength.
Close to the $A'$ resonance we find $\epsilon\sim10^{-4}$–$10^{-3}$, while near the $Z$ resonance we find $\epsilon\sim10^{-1}$. 
The smaller values near $m_{A'}/2$ reflect the much narrower width, $\Gamma_{A'}\ll \Gamma_Z$, which makes the Breit–Wigner enhancement narrower and taller for $A'$.

For $g_X=0.1$ and $m_{A'}=200~\mathrm{GeV}$ the correct relic density is obtained over a broad range, $m_\chi\simeq 20$–$400~\mathrm{GeV}$. 
Increasing $m_{A'}$ or lowering $g_X$ squeezes the viable region to narrow windows near the resonances: the neighborhood of $m_Z/2$ becomes very thin, while around $m_{A'}/2$ one typically needs $m_\chi$ within $\sim 10$–$20\%$ of the resonance (the precise range depends on $g_X$).

Including the LZ constraint excludes most of the $(m_\chi,\epsilon)$ plane. 
The robustly surviving region for all $(m_{A'},g_X)$ choices is the $A'$–resonance funnel, $m_\chi\simeq m_{A'}/2$, where a tuning of
\begin{equation}
0\;\lesssim\;\frac{m_{A'}-2m_\chi}{m_{A'}}\;\lesssim\;\xi
\end{equation}
is required, with $\xi\sim 1\%$ for $g_X=0.1$ and up to $\xi\sim 6\%$ for $g_X=0.01$ in our scan.

We also overlay the \emph{Fermi}–LAT dSphs bounds~\cite{McDaniel:2023bju}. 
Their shape tracks the relic curve because the present–day annihilation cross section is likewise resonantly enhanced. 
Away from $m_{A'}/2$ and $m_Z/2$ the indirect–detection limits are much weaker than LZ, whereas very close to the funnels they can become competitive and may exclude the most resonant points.

\subsubsection{Secluded regime}

\begin{figure}[t]
  \centering
  \includegraphics[width=0.98\linewidth]{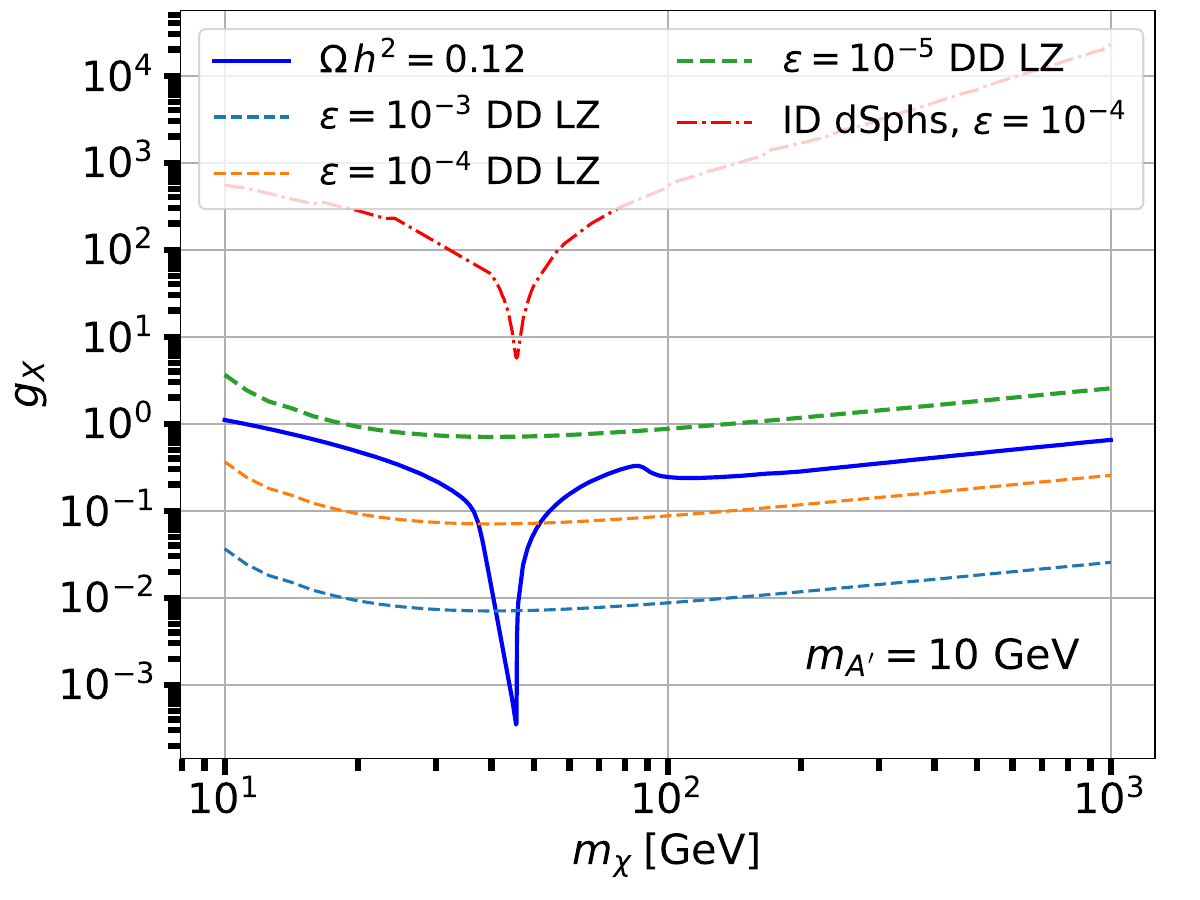}
  \caption{We show in this figure the \emph{secluded} case for Model I with $m_A'=10$ GeV. We report the parameters $g_X$ and $m_\chi$ for which DM has the right relic density (blue solid curve) and the upper limits obtained when fixing different values of $\epsilon$ from $10^{-5}$ and $10^{-3}$. Finally we report the upper limits for $g_X$ obtained when fixing $\epsilon=10^{-4}$ from indirect detection (red dot-dashed curve).}
\label{fig:ModIsecluded}
\end{figure}

In the secluded regime we take $m_{A'}=\{1,\,10\}\,\mathrm{GeV}$ and proceed as in the previous section.
Figure~\ref{fig:ModIsecluded} shows the case $m_{A'}=10~\mathrm{GeV}$; the $m_{A'}=1~\mathrm{GeV}$ case is qualitatively similar.
We plot in the $(m_\chi,g_X)$ plane the parameter values that reproduce the observed relic abundance, together with direct–detection limits recast as upper bounds on $g_X$ for fixed benchmark portal couplings $\epsilon=10^{-3},\,10^{-4},\,10^{-5}$.
In the kinetic–mixing model, $\sigma_{\chi N}^{\rm SI}\propto \epsilon^2 g_X^2/m_{A'}^{4}$, hence a bound on $\sigma_{\chi N}^{\rm SI}$ translates into a bound on $g_X$ for each chosen $\epsilon$.

As expected for the secluded case $m_\chi>m_{A'}$, the relic density is dominantly set by $\chi\bar\chi\to A'A'$, so the blue curve is essentially $g_X(m_\chi)$ fixed by the thermal target and is only weakly sensitive to $\epsilon$.
Direct–detection limits scale as $\sigma_{\rm SI}\propto \epsilon^{2}g_X^{2}/m_{A'}^{4}$; for each benchmark $\epsilon$ the LZ bound therefore translates into an upper limit on $g_X$ (dashed curves), tighter for larger $\epsilon$.

For $\epsilon\gtrsim 10^{-4}$ the direct-detection bound excludes most masses except for narrow resonant regions (notably near $m_\chi\simeq m_Z/2$, where the $s$–channel $\chi\bar\chi\to f\bar f$ rate is resonantly enhanced and the required $g_X$ along the relic curve is reduced).
For $\epsilon\lesssim 10^{-4}$ essentially the entire mass range above $10~\mathrm{GeV}$ remains viable without relying on fine resonance tuning, since the direct-detection limits relax as $\propto \epsilon^{-1}$ at fixed $m_{A'}$.
Indirect–detection limits from dSphs are too weak in this $\epsilon\lesssim 10^{-4}$ regime to constrain the parameter space shown; the present–day cross section into SM states is suppressed by the same small portal.

In summary, for $10^{-7}\!\lesssim\!\epsilon\!\lesssim\!10^{-4}$ and $m_{A'}\!\lesssim\!10~\mathrm{GeV}$, the observed relic abundance can be obtained with
$g_X=\mathcal{O}(0.1\text{--}1)$ over a wide range of $m_\chi\gtrsim 10~\mathrm{GeV}$ without sitting on a resonance.
The lower end of this $\epsilon$ range also satisfies the requirement that $A'$ remain in chemical contact with the SM at $T\sim m_{A'}$ and that its decays occur before BBN (cf.\ Sec.~\ref{sec:BBN}).

\subsubsection{Sommerfeld enhancement and bound-state formation}
\label{sec:SEBS}

If two slowly moving DM particles feel an \emph{attractive, long-range} force (from a light mediator), their incoming two-body wavefunction is distorted and piles up at short distances, boosting the short-distance annihilation rate by a multiplicative factor that for $s$-wave annihilation can be written as
\begin{equation}
\langle \sigma v_{\rm{rel}} \rangle_{\text{eff}} \,=\, S(v)\,\langle \sigma v_{\rm{rel}} \rangle_0,
\qquad 
S \equiv \frac{|\psi(0)|^2}{|\psi_0(0)|^2}.
\label{eq:defS}
\end{equation}
Here $\langle \sigma v_{\rm{rel}} \rangle_0$ is the tree-level (short-distance) annihilation, while $S$ encodes long-range distortion \cite{Cassel:2009wt,Feng:2010zp}.

If the attraction is strong/long-ranged enough to support hydrogen-like levels, a DM pair can \emph{radiatively capture} into a bound state by emitting a mediator (or other light quanta). The bound state typically decays (annihilates) promptly, providing an additional depletion channel during freeze-out and an extra source for indirect signals today \cite{An:2016gad,Mitridate:2017izz,vonHarling:2014kha}.
Ref.~\cite{Pospelov:2008jd} reported that Sommerfeld enhancement and bound-state formation can be relevant for secluded DM models and for MeV scale mediator.

When $m_\phi \gtrsim m_\chi v$ it is customary to consider an attractive central Yukawa potential
\begin{equation}
V(r) \;=\; -\,\frac{\alpha_{\rm eff}}{r}\,e^{-m_\phi r},
\label{eq:Yukawa}
\end{equation}
with mediator mass $m_\phi$, coupling $\alpha_{\rm eff}$ in the \emph{attractive} annihilating channel, DM mass $m_\chi$, and relative velocity $v$.
It is useful to define the dimensionless ratios
\begin{equation}
\epsilon_\phi \;\equiv\; \frac{m_\phi}{\alpha_{\rm eff}\,m_\chi},
\qquad
\epsilon_v \;\equiv\; \frac{v}{2\,\alpha_{\rm eff}}.
\label{eq:epsdefs}
\end{equation}
Three length scales control the behavior: the force range $r_\phi=1/m_\phi$, the Bohr radius $a_0\simeq 1/(\mu\alpha_{\rm eff})\simeq 2/(\alpha_{\rm eff}m_\chi)$, and the de Broglie wavelength $\lambda_{\rm dB}\simeq 2/(m_\chi v)$.
A \emph{necessary} long-range condition for sizable Sommerfeld enhancement is that the force range is larger than the Bohr radius $r_\phi>a_0$ that implies
\begin{equation}
\epsilon_\phi \;=\; \frac{m_\phi}{\alpha_{\rm eff}m_\chi}\;\lesssim\;\mathcal{O}(1).
\label{eq:longrange}
\end{equation}
If $\epsilon_\phi\gtrsim 1$ the interaction is effectively short-range and $S\simeq 1$.
Near the appearance of zero-energy bound states, $S(v)$ develops narrow \emph{resonances}; in the Hulth\'en approximation the condition is roughly \cite{Feng:2010zp}
\begin{equation}
\frac{\alpha_{\rm eff}m_\chi}{m_\phi} \;\approx\; \frac{\pi^2 n^2}{6}, \qquad n=1,2,\dots
\label{eq:resonances}
\end{equation}
A Yukawa potential is considered when the DM de Broglie wavelength $\lambda_{\rm dB}$ is larger than $r_\phi$. In the opposite regime instead a Coulomb potential is assumed.
In the Yukawa case (Coulomb) the condition $r_\phi \lesssim \lambda_{\rm dB}$ ($r_\phi \gg \lambda_{\rm dB}$) implies $m_\phi \gtrsim m_\chi  v$ ($m_\phi \ll m_\chi  v$) so the mediator can be considered as massive (massless).

In the $U(1)_X$ model reported in this section with DM particle $\chi$ and mediator $A'$, the $\chi\bar\chi \to A'A'$ annihilation channel is attractive and the previous equations apply with the substitutions $\alpha_{\rm eff}\to \alpha_X\equiv g_X^2/4\pi$ and $m_\phi\to m_{A'}$.
At $m_\chi=100$ GeV with $\alpha_X$ fixed by $\chi\bar\chi\to A'A'$ to the observed relic density, Sommerfeld enhancement is relevant only for $m_{A'}\lesssim{\cal O}(0.1$--$0.3)$ GeV. Instead, bound-state formation via single $A'$ emission is kinematically closed unless $m_{A'}\!\lesssim\!0.15~$ MeV.
These mass values for $A'$ are consistent with what found in Ref.~\cite{Pospelov:2008jd}. Since in our paper we consider GeV scale mediators we do not take into account Sommerfeld enhancement and bound-state formation.

\subsection{Model II ({\tt DMSimp} with scalar DM and scalar mediator)}

\begin{figure*}[t]
  \centering
  \includegraphics[width=0.49\linewidth]{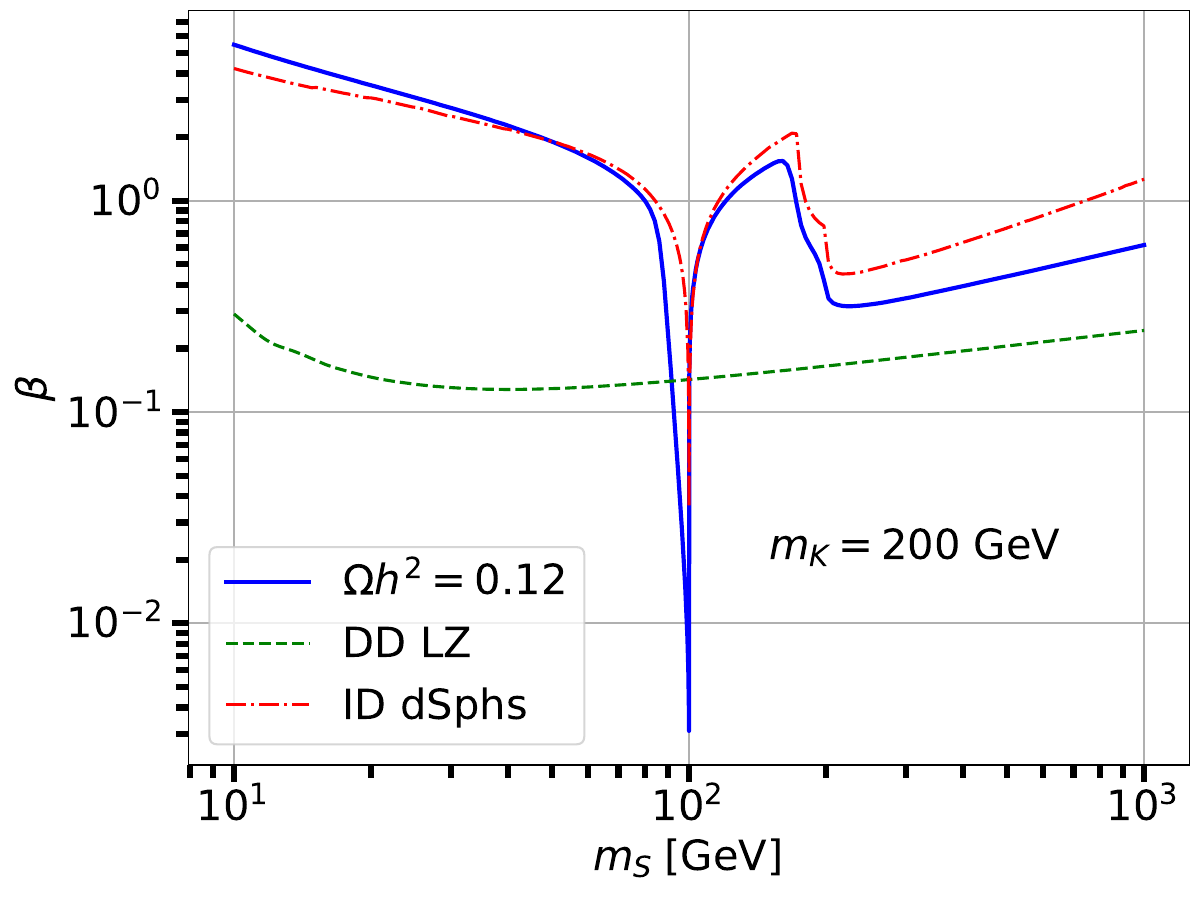}
  \includegraphics[width=0.49\linewidth]{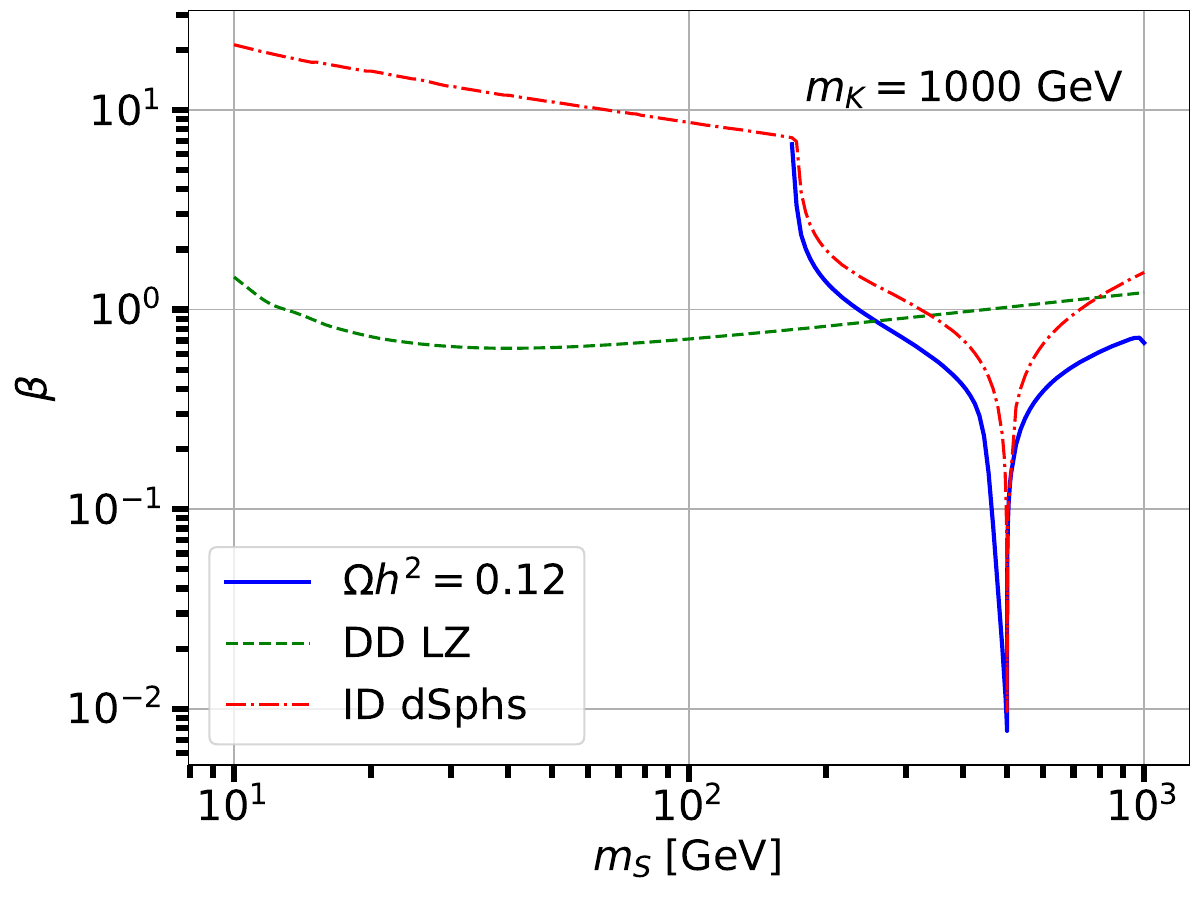}
  \includegraphics[width=0.49\linewidth]{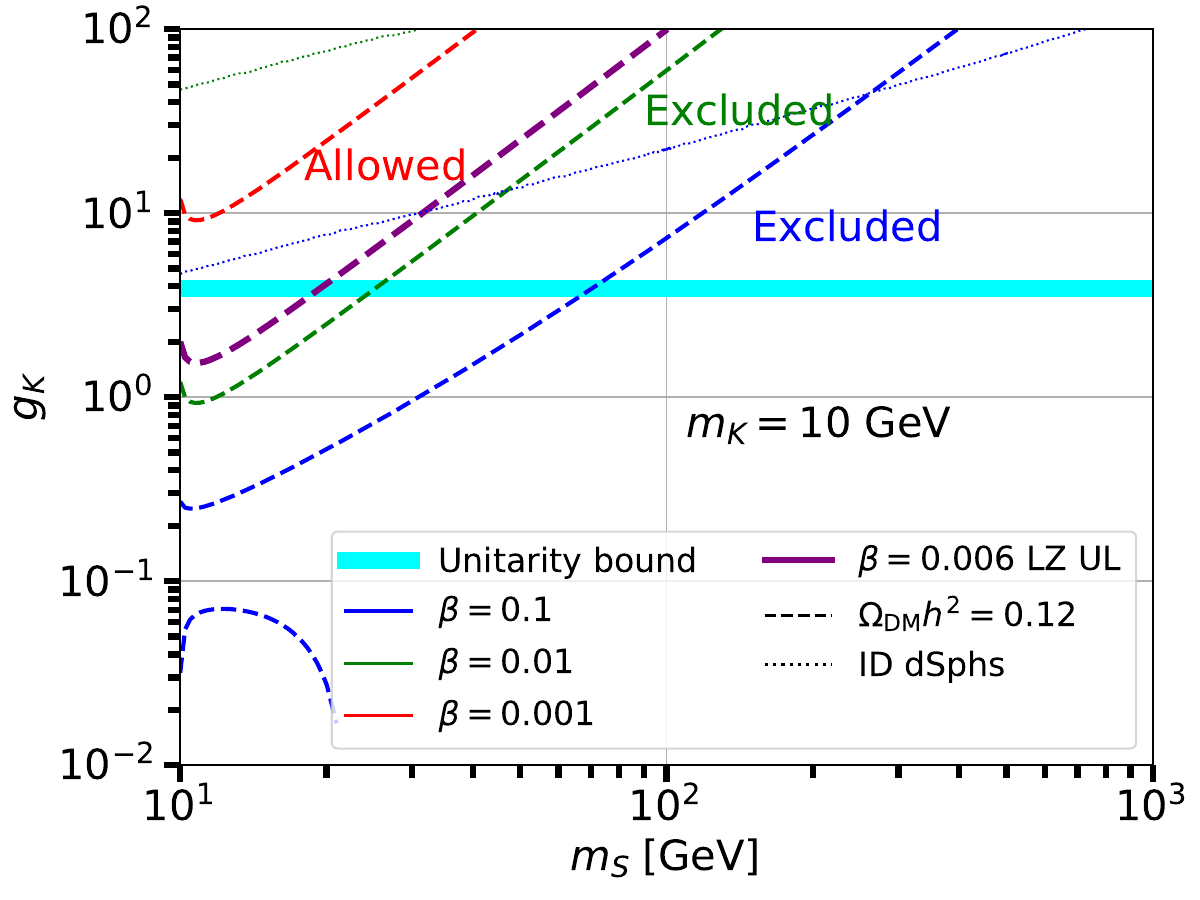}
  \includegraphics[width=0.49\linewidth]{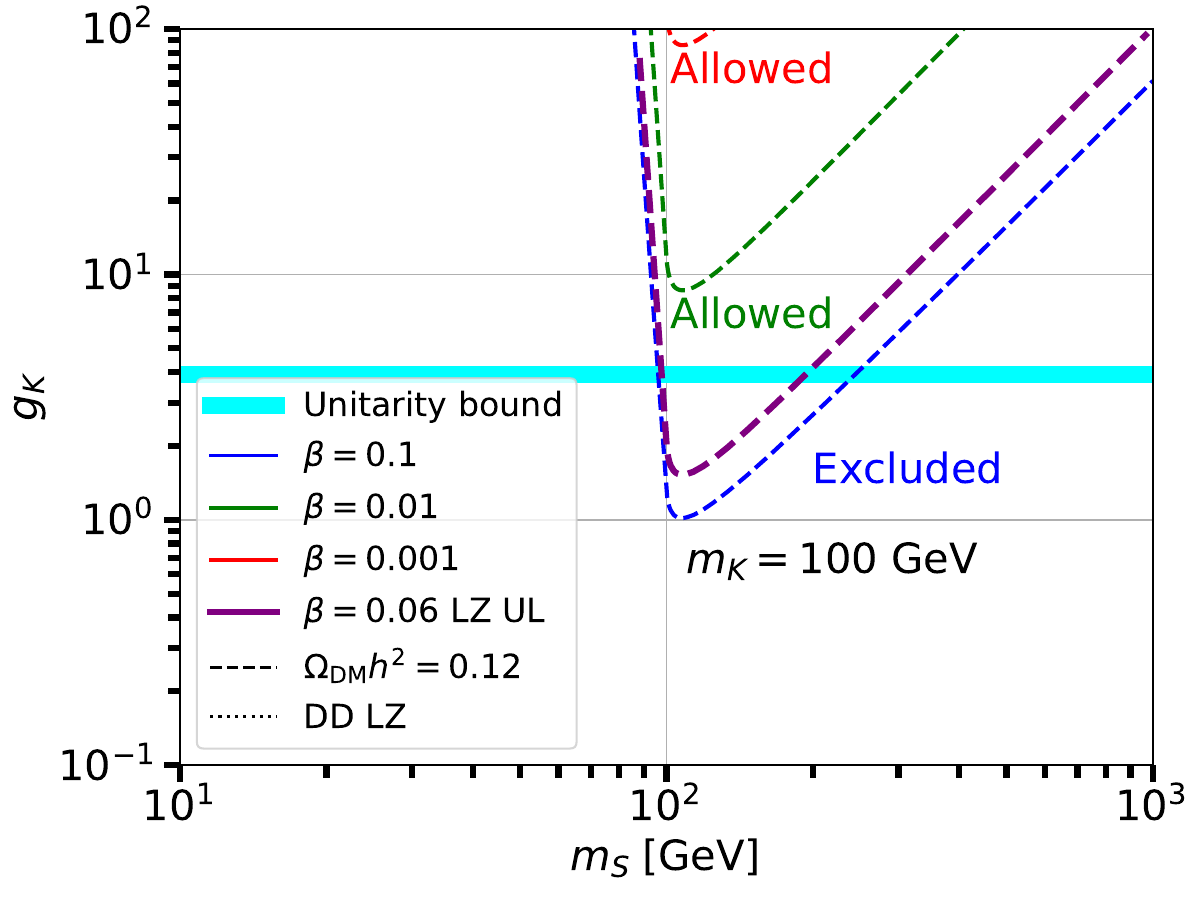}
  \caption{This figure shows the results obtained for the Model II. The top panels include the WIMP scenario with the values $m_S$ and $\beta$ for which DM has the right relic density (blue solid) compared with the upper limits obtained from LZ (green dashed) and dSphs bounds (red dot-dashed). The left (right) panel display the case with $m_K=200$ (1000) GeV. The bottom figures show the secluded case with $m_K=10$ GeV (100 GeV) in the left (right) panel. We show with a cyan band the unitarity bound for the coupling parameter $g_K$, the curves with values of DM mass and $g_K$ that provides the correct relic density (dashed curves) obtained for $\beta=[10^{-3}-10^{-1}]$. Next to each relic density curve we report if these model parameters are excluded or allowed by LZ bounds.}
\label{fig:DMSimp}
\end{figure*}

\subsubsection{WIMP scenario}

For the second model we follow the same strategy as for Model~I. 
We first consider the \emph{WIMP regime}, in which the relic density is predominantly achieved through DM annihilation into SM fermions. The parameter $\beta$ controls the Yukawa–like coupling of the mediator $K$ to SM fermions.
The coupling $\lambda_{KS}$, which controls the interaction terms $S^2K^2$ and $S^2K$, is fixed to be equal to $\beta$.

We show in Fig.~\ref{fig:DMSimp} two benchmarks,
$m_K=200~\mathrm{GeV}$ and $m_K=1000~\mathrm{GeV}$. 
In each panel we determine the value of the mediator–fermion coupling $\beta$ that reproduces the observed relic density, and we compare it with the direct–detection (LZ) upper limits and with indirect–detection bounds from dSphs.
Away from the $s$–channel resonance at $m_S\simeq m_K/2$, the required coupling lies in the range 
\(\beta\simeq 0.3\)–\(5\), depending on $m_S$. 
As $m_S$ approaches $m_K/2$, the annihilation cross section
$SS\!\to\! f\bar f$ is resonantly enhanced and the coupling needed to achieve 
$\Omega_{\rm DM}h^2\simeq0.12$ drops sharply to $\beta\sim 10^{-3}$–$10^{-2}$.

Direct detection excludes a large fraction of the $(m_S,\beta)$ plane. 
Since the scalar–mediated SI cross section scales as $\sigma_{\rm SI}\propto (\beta\,\lambda_{KS})^{2}/m_K^{4}$, the LZ bound translates into an upper limit on $\beta$ that weakens as $m_K$ increases. 
For $m_K=200~\mathrm{GeV}$ only a very narrow region around the resonance survives the LZ constraint (the DM mass must lie within a few percent of $m_K/2$ in our scan). 
For $m_K=1000~\mathrm{GeV}$ a broader allowed region appears on the high–mass side of the resonance; in our scan the relic curve falls below the LZ limit for $m_S$ well above the resonance (several hundred GeV).

The dSph limits track the shape of the relic curve, both are enhanced on resonance, but are otherwise much weaker than direct detection and become relevant only very close to $m_K/2$.

In summary, for mediator masses below the TeV scale the WIMP region typically requires a degree of tuning near $m_K/2$ to satisfy LZ, whereas for $m_K\simeq 1~\mathrm{TeV}$ the simultaneously allowed region widens (though it still clusters around the resonance).

\subsubsection{Secluded regime}

We now consider the secluded case, with benchmarks $m_K=\{1,\,10,\,100\}\,\mathrm{GeV}$ and representative values $\beta=\{10^{-1},10^{-2},10^{-3}\}$. 
The dominant annihilation process setting the relic density is $S S\to K K$, which receives: (i) contact $S^2K^2$ and $t/u$–channel $S$ contributions $\propto \lambda_{KS}^4$, and
(ii) an $s$–channel $K$ contribution, controlled by the interaction term $g_K \,m_K\,K^3$, $\propto \lambda_{KS}^2 g_K^2$. Near threshold the latter often dominates.
As in the previous section, the quartic portal $\lambda_{KS}$ is kept equal to $\beta$.

We show in Fig.~\ref{fig:DMSimp} the relic density curve in the $(m_S,g_K)$ plane.
For fixed $(m_K,\beta)$ the dashed curves show the values of $g_K(m_S)$ that reproduce $\Omega_{\rm DM}h^2\simeq 0.12$.
Next to them we annotate whether the corresponding spin–independent
cross section is excluded or allowed by the LZ limit.
Since the scalar–mediated nucleon cross section scales as
$\sigma^{\rm SI}_{SN}\ \propto\ (\beta\,\lambda_{KS})^2/m_K^4$,
light mediators are strongly constrained: for $m_K=1~\mathrm{GeV}$ one needs $\beta\lesssim 10^{-3}$ to evade LZ, but then achieving the relic density along the dashed curve forces $g_K\gtrsim 10^3$, which is far beyond the perturbative/unitarity regime.

For heavier mediators the direct–detection bound weakens.
At $m_K=10~(100)\,\mathrm{GeV}$, the minimum $\beta$ compatible with LZ
in our setup is $\beta_{\rm min}\simeq 6\times 10^{-3}\ (6\times 10^{-2})$.
With these values, the relic curve requires $g_K\gtrsim 2$ over the mass
interval shown.

Since we need values $g_K\gtrsim 2$ to reach the relic density and being compatible with direct detection bounds we have to check if these values violates the unitarity of the theory.
The most constraining perturbative unitarity limit on $g_K$ comes from elastic $K K\to K K$ scattering via the cubic vertex $g_K m_K K^3$. The tree–level amplitude is
\begin{equation}
\mathcal{M}(s,t,u)=
\mu^2\!\left(\frac{1}{s-m_K^2}+\frac{1}{t-m_K^2}+\frac{1}{u-m_K^2}\right),
\end{equation}
which at threshold ($s=4m_K^2$, $t=u\to 0$) gives
\begin{equation}
\mathcal{M}_{\rm th}
= g_K^2 m_K^2\!\left(\frac{1}{3m_K^2}-\frac{1}{m_K^2}-\frac{1}{m_K^2}\right)
= -\frac{5}{3}\,g_K^2.
\end{equation}
The $s$–wave partial amplitude is
$a_0=\mathcal{M}_{\rm th}/(16\pi)$, hence imposing
$|\mathrm{Re}\,a_0|\le \tfrac12$ yields
\begin{equation}
|g_K|\ \le\ \sqrt{\frac{24\pi}{5}}\ \simeq\ 3.9\,,
\end{equation}
a robust, mass–independent cap. The cyan band in Fig.~\ref{fig:DMSimp}
shows this limit.

Combining relic density, LZ, and unitarity, we find that for
$m_K=10~\mathrm{GeV}$ ($100~\mathrm{GeV}$) a viable region exists with $g_K \in [1.5,\,3.9]$, $m_S \in [10,\,20]~\mathrm{GeV}$ ($[100,\,200]$ GeV), while the $m_K=1~\mathrm{GeV}$ case is excluded by the tension between
the LZ bound (which drives $\beta$ very small) and the unitarity bound
on the large $g_K$ then required by relic density.
Indirect–detection limits (dotted) track the relic curve but are otherwise subdominant to LZ in the mass ranges shown.

\subsection{Model III ($U(1)_X$, with Dirac DM and Higgs-mixed singlet)}

In this section we report the results obtained for Model III where the DM particle is a Dirac fermion $\chi$ and there is a new scalar $H_p$ which is the one that provides mass to $\chi$ through a Yukawa term $y_p \bar{\chi} \chi R$.
We fix $m_Z'=1$ TeV, $g_X=1$ and $\lambda_{HR}=0.1$. These parameters are not that relevant for the phenomenology of the model.

\begin{figure}
  \centering
  \includegraphics[width=0.99\linewidth]{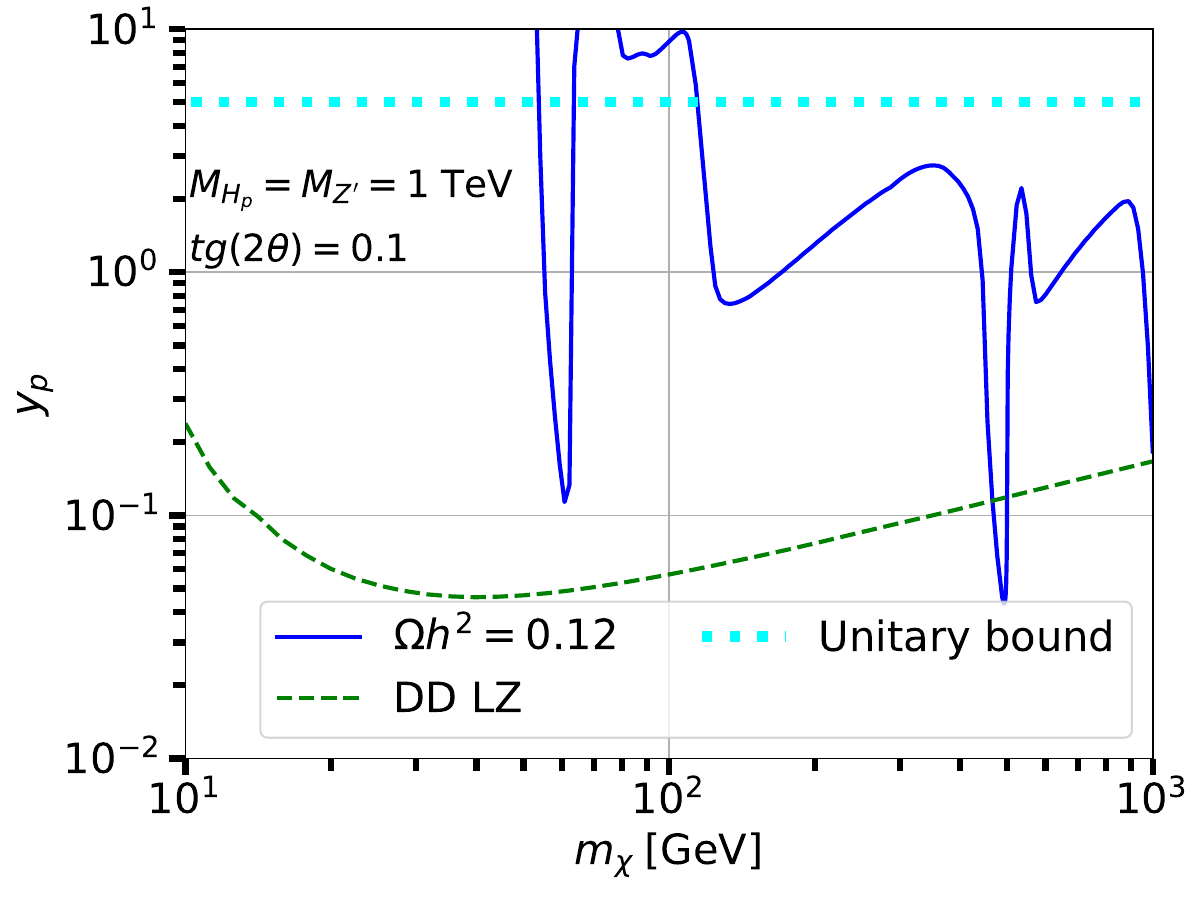}
  \includegraphics[width=0.99\linewidth]{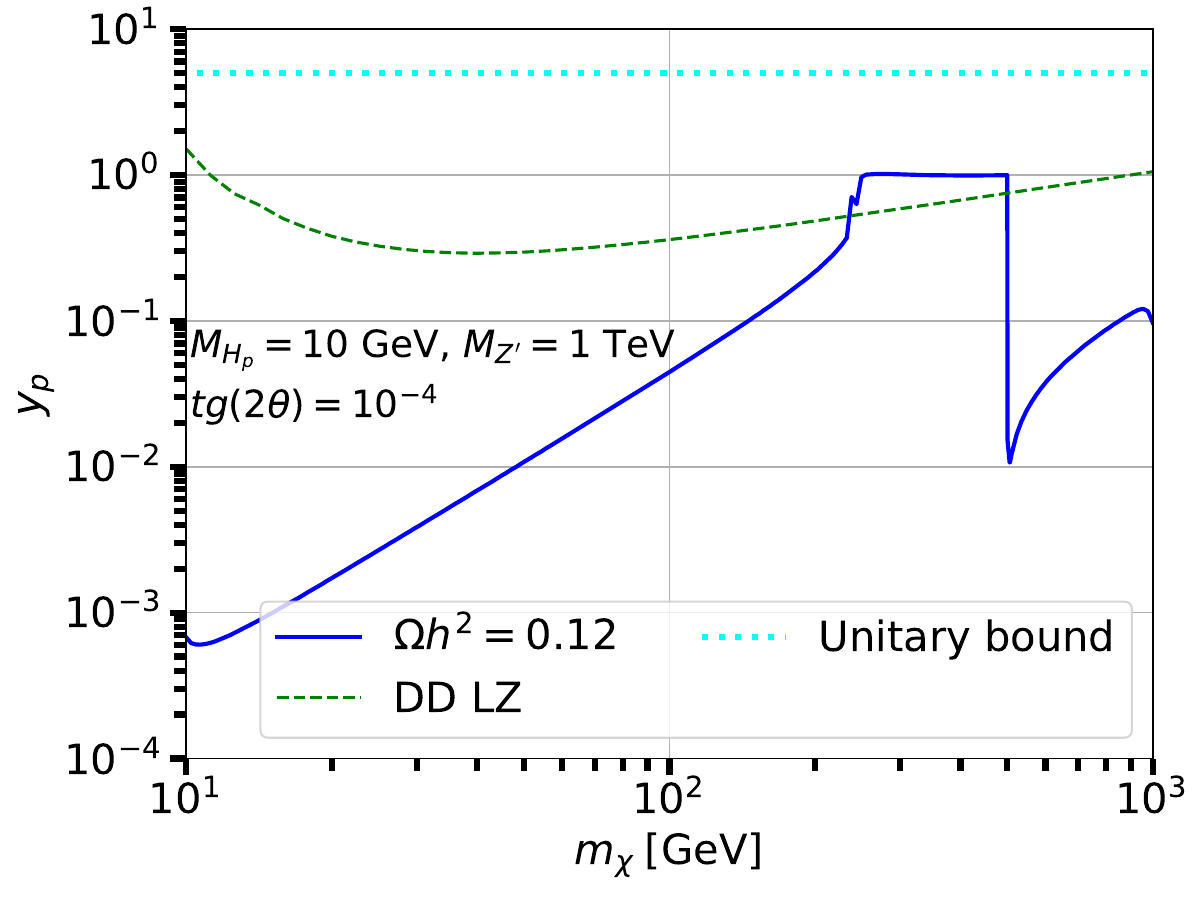}
  \includegraphics[width=0.99\linewidth]{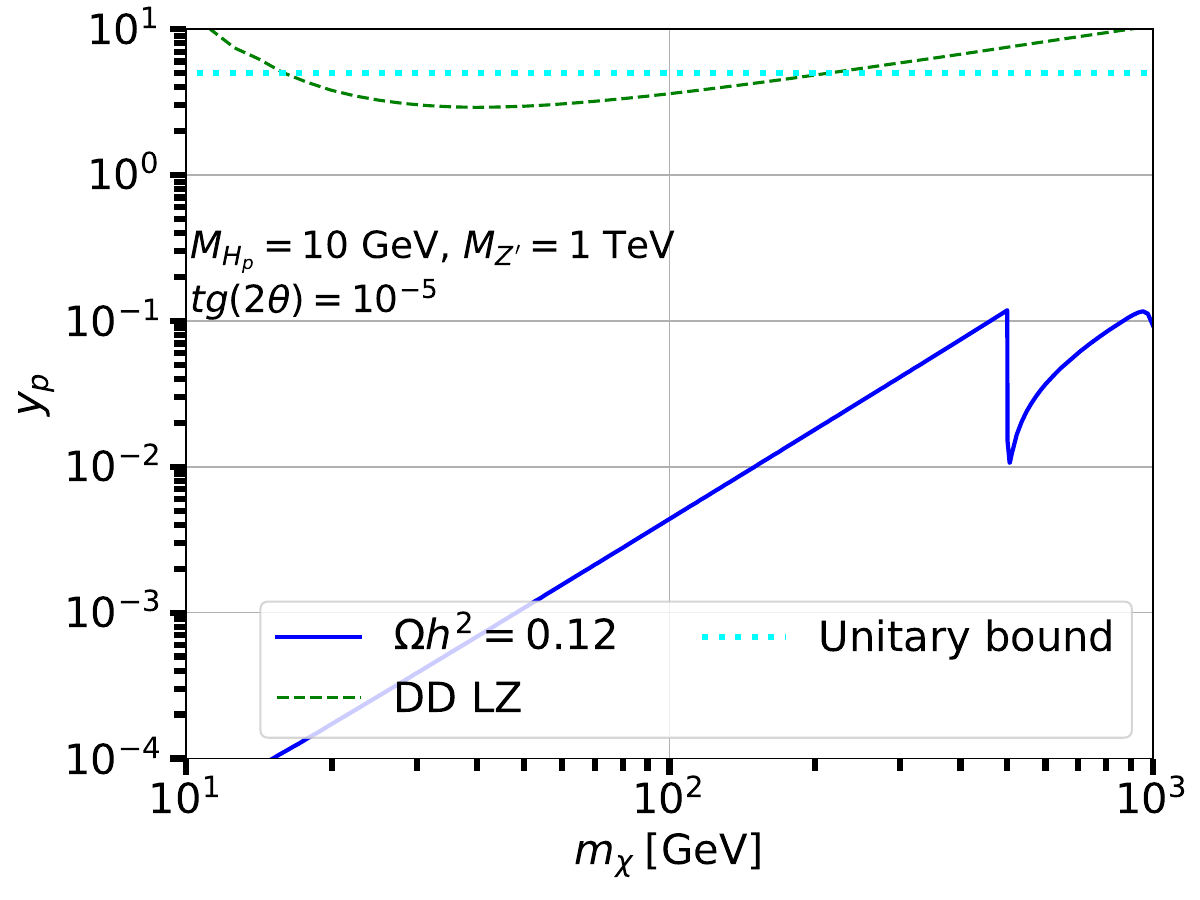}
  \caption{The top panel shows the values of the parameters $m_\chi$ and $y_p$ that provide the correct value of the DM relic density (blue solid curve) compared with the upper limits obtained from LZ nuclear cross section. We have fixed $\tan{2\alpha}=0.1$ and $m_{H_p}=1$ TeV that represents the WIMP regime for the model for which DM annihilates predominantly into SM fermions.
  In the central and bottom panel we show the same result when we fix $m_{H_p}=1$ TeV and $\tan{2\alpha}=10^{-4}$ and $10^{-5}$. For these choices of the model parameters we are in the secluded regime where DM annihilates mostly into $H_p$ pairs.}
\label{fig:DMHiggs}
\end{figure}

\subsubsection{WIMP scenario}

In the WIMP scenario the most relevant process that brings to the DM relic density is the annihilation into SM fermion pairs through the Higgs-like particles $H$ and $H_p$ with a cross section proportional to $y_p^2 (\sin\alpha\cos\alpha)^2$ (see Eq.~\ref{eq:sigmaff-scalar}).
For small values of the couplings ($\alpha\ll\pi$) $\sin\alpha\cos\alpha\sim \tan{2\alpha}/2$ so the annihilation cross section into SM fermions is proportional to $y_p^2 (\tan{2\alpha})^2$.

In top panel of Fig.~\ref{fig:DMHiggs} we show the outcome of the analysis when we fix the $H_p$ mass to 1 TeV and $\tan{(2\alpha)}=0.1$.
%In this case the DM annihilation process that dominates the relic density is $\chi \bar{\chi} \to f \bar{f}$ mediated by $S$. 
%If the mixing angle $\alpha$ between the two Higgs-like particles is very small the annihilation process is mediated by the new scalar $H_p$.
We report the values of the parameter space $m_\chi-y_p$ that explain the observed value of the relic density.
Values of $y_p\sim \mathcal{O}(0.1-1)$ are needed away from $m_H/2$ and $m_{H_p}/2$.
Instead, approaching the resonant regions much smaller values of $y_p$ of the order of $y_p\sim \mathcal{O}(0.1)$
are needed to fit $\Omega_{\rm{DM}}h^2=0.12$.
The upper limits on the spin-independent direct detection are very strong and rule out most of the parameter space, except for the resonant region around $m_{H_p}/2$ with a needed fine tune of the DM and $H_p$ masses of the order of $10\%$.

When using values of $\tan{(2\alpha)}$ smaller than 0.1 the relic density curve is even more confined towards the resonant region around $H_p$ and for $\tan{(2\alpha)}<10^{-3}$ it is not possible to reach the correct DM relic density.

\subsubsection{Secluded case}

In the secluded case and for small mixing angles $\alpha$, the DM relic density is set predominantly by annihilations into $H_p$ and $Z'$ pairs. 
We consider $m_{H_p}=10$~GeV and values of $\tan(2\alpha)$ between $10^{-5}$ and $10^{-1}$.
For this $H_p$ mass, the processes that dominate the relic density for DM masses up to about $m_{Z'}/2$ are the $t$- and $u$-channel annihilations into $H_p H_p$, which scale as $y_p^4 \cos^4\!\alpha$ (see Eq.~\ref{eq:sigmaHH}). 
Therefore, when $\tan(2\alpha)\ll 1$, $\cos\alpha\simeq 1$ and the relic density is controlled primarily by $y_p$.

By contrast, the spin-independent nuclear cross section receives two contributions—from the SM Higgs and from $H_p$—and is proportional to $y_p^2 (\sin\alpha\,\cos\alpha)^2 \simeq y_p^2 \big(\tan(2\alpha)/2\big)^2$ for small mixing angles.
Consequently, the LZ constraints on $y_p$ become weaker as $\tan(2\alpha)$ decreases.

The central and bottom panels of Fig.~\ref{fig:DMHiggs} show our results.
For $\tan(2\alpha)>10^{-3}$, most of the parameter space yielding the correct relic density is excluded by LZ.
For smaller values, the direct-detection limits weaken significantly, scaling as $\tan^2(2\alpha)$. In particular, reducing $\tan(2\alpha)$ by a factor of 10 relaxes the upper limit on $y_p$ by the same factor.
This implies that for $\tan(2\alpha)<10^{-4}$ most of the parameter space is allowed by the LZ bounds.

Dark matter can attain the observed relic density across the full mass range considered in our analysis.
In particular, the coupling $y_p$ required to obtain $\Omega_{\rm DM}h^2$ increases with $m_\chi$. This is because the annihilation cross section into $H_p$ pairs decreases with $m_\chi$ (see Eqs.~\ref{eq:sigmaHH} and \ref{eq:sigma_schannel_NR}). Therefore, as $m_\chi$ grows, $\langle \sigma v_{\rm{rel}} \rangle_{H_pH_p}$ becomes smaller and a larger $y_p$ is needed to compensate. Empirically, the trend that reproduces the correct relic density is approximately $y_p \propto m_\chi^2$, which lies between the scaling from the $t$- and $u$-channels ($y_p\propto \sqrt{m_\chi}$; see Eq.~\ref{eq:sigmaHH}) and that from the $s$-channel ($y_p\propto m_\chi^3$; see Eq.~\ref{eq:sigma_schannel_NR}).

To conclude, in Model~III one can evade direct-detection bounds by choosing a very small mixing angle, i.e., $\tan(2\alpha)<10^{-4}$. For such $\alpha$, the main process setting the relic density is $\chi\bar\chi\to H_p H_p$, with
\begin{equation}
y_p \sim 5\times 10^{-4}\left(\frac{\tan(2\alpha)}{10^{-4}}\right) \left( \frac{m_\chi}{10\,\mathrm{GeV}}\right)^2,
\end{equation}
for $m_\chi \lesssim m_{Z'}/2$, above which annihilation into $Z'Z'$ becomes relevant.

\section{Possible Experimental Probes for Secluded Dark Matter}
\label{sec:tests}

In this section we discuss experimental tests for secluded DM beyond the direct- and indirect-detection strategies analyzed above. We provide a qualitative overview of collider, astrophysical, and cosmological probes, and defer to follow-up work for a quantitative treatment.

\subsection{Collider searches for secluded dark matter}

In secluded scenarios the direct-detection signal is often too weak for present and future experiments because the portal is tiny, yet several laboratory and cosmological handles remain powerful. The key point is that direct detection and colliders scale differently with the portal couplings, and their backgrounds and kinematics are distinct.

For spin–independent direct detection the (schematic) event rate scales as
\begin{equation}
R_{\rm DD}\;\propto\;\frac{\rho_\chi}{m_\chi}\;\sigma_{\chi N}^{\rm SI}
\;\sim\;\frac{\rho_\chi}{m_\chi}\;
\frac{\big(g_{X}\,\epsilon\big)^2}{m_{\phi}^4},
\end{equation}
so a heavy mediator (off-shell exchange $\propto 1/m_{\phi}^4$) and the portal entering \emph{quadratically} drive $\sigma_{\chi N}^{\rm SI}$ below the neutrino floor once $\epsilon\ll 10^{-3}$ (for a vector portal; analogous suppressions apply in the other models).

By contrast, at colliders and intensity-frontier experiments, mediators are produced \emph{on shell}: the off-shell $1/m_{\phi}^4$ suppression is replaced by a Breit–Wigner propagator, yielding a resonant enhancement near the pole $s\simeq m_{\phi}^2$,
\begin{equation}
\sigma_{\rm coll}\ \propto\ 
\epsilon^2\,
\frac{m_{\phi}\,\Gamma_{\phi}}{(s-m_{\phi}^2)^2+m_{\phi}^2\Gamma_{\phi}^2}\;
{\rm BR}(\phi\!\to\! X).
\label{eq:BW}
\end{equation}
In the narrow–width approximation the observable (visible) signal factorizes,
\begin{equation}
\sigma_{\rm vis}\ \simeq\ \sigma_{\rm prod}\times {\rm BR}_{\rm vis}
\ \propto\ \epsilon^2\times {\rm BR}_{\rm vis},
\label{eq:NWA}
\end{equation}
where ${\rm BR}_{\rm vis}\equiv \Gamma_{\phi,{\rm vis}}/\Gamma_{\phi}$. On–shell production removes the $1/m_{\phi}^4$ penalty and brings a resonant enhancement; the \emph{integrated} yield still scales as $\epsilon^2$ times the relevant branching ratio. At $s=m_{\phi}^2$ one has $\sigma_{\rm vis}\propto \epsilon^2/(m_{\phi}\Gamma_{\phi})$; if $\Gamma_{\phi}\propto \epsilon^2$ (visible-width dominated) the peak yield becomes approximately portal–independent, while if a dark width dominates it scales as $\epsilon^2/\Gamma_{\rm tot}$.

In secluded models with tiny portals, present–day annihilations in halos occur at $v\sim10^{-3}$, so $s\simeq 4m_\chi^2(1+v^2/4)$ is effectively fixed. To benefit from an $s$–channel pole one must satisfy
\[
\left|\frac{m_{\phi}^2-4m_\chi^2}{m_{\phi}^2}\right|\;\lesssim\;\max\!\left(\frac{\Gamma_{\phi}}{m_{\phi}},\,\frac{v^2}{4}\right).
\]
With tiny portals the mediator is ultra–narrow, $\Gamma_{\phi}/m_{\phi}\propto \epsilon^2\!\ll\!1$ (e.g.\ $10^{-10}$–$10^{-12}$ for $\epsilon\sim10^{-4}$–$10^{-5}$), while $v^2/4\sim2.5\times10^{-7}$. Unless $2m_\chi$ is tuned to the pole at the $10^{-7}$–$10^{-10}$ level, today’s annihilation is \emph{off} resonance and the rate collapses to its suppressed value. In addition, many secluded leading channels are $p$–wave ($\langle \sigma v_{\rm{rel}} \rangle\propto v^2$), giving another $\sim10^{-6}$ suppression relative to freeze-out. Indirect searches must also contend with $J$–factor uncertainties and astrophysical backgrounds. By contrast, colliders and intensity–frontier experiments can produce the mediator on shell, effectively \emph{scan the pole} via center–of–mass energy or parton luminosities, and exploit ultra–low–background signatures (e.g.\ displaced decays).

\subsection{Experimental handles beyond direct detection}
\label{sec:exp_handles}

Even when the portal to the SM is tiny and spin–independent scattering is far below current or next–generation sensitivities, several laboratory strategies remain incisive.

\emph{Long–lived particle (LLP) searches.}
Small portals naturally yield long mediator lifetimes. Displaced dileptons or lepton–jets at ATLAS, CMS, and LHCb, as well as dedicated forward detectors such as FASER and proposed surface/forward facilities (e.g.\ MATHUSLA), operate in ultra–low–background regimes and retain excellent sensitivity across the characteristic “lifetime window’’ where particles decay within instrumented volumes \cite{Alimena:2019zri,FASER:2018bac,Curtin:2018mvb}.

\emph{Exotic Higgs decays.}
If the mediator (or Higgs–mixed scalar) is lighter than half the Higgs mass, exotic decays like $h\to\phi\phi\to 4f$ become powerful probes, with prompt or displaced topologies depending on the portal. These channels directly constrain Higgs–portal couplings and scalar mixing angles, independently of the local DM density, and will be pushed further by the HL–LHC and future lepton colliders \cite{Curtin:2013fra}.

\emph{Intensity–frontier experiments.}
Fixed–target and beam–dump facilities (APEX, HPS, LDMX, NA64, SHiP), as well as high–luminosity $e^+e^-$ factories (Belle~II), produce mediators on shell with clean visible or invisible final states and modest backgrounds. Their enormous effective luminosities allow them to probe portal couplings far below what is accessible at hadron colliders or direct detection \cite{Abrahamyan:2011gv,HPS:2019pow,Akesson:2018vlm,NA64:2017vtt,Anelli:2015pba,Kou:2018nap}.

\emph{Precision Higgs and electroweak fits.}
In models with scalar mixing, Higgs signal strengths are universally rescaled, enabling global fits to set robust bounds on the mixing angle regardless of nuclear matrix elements or astrophysical uncertainties. Current and projected fits already probe parameter space that is otherwise invisible to direct searches \cite{Robens:2015gla,Cepeda:2019klc}.

\emph{Astrophysical and cosmological probes.}
Light mediators can induce self–interactions that alter small–scale structure (dwarfs, clusters, strong–lensing substructure), providing tests that are essentially independent of the SM portal \cite{Tulin:2017ara,Vegetti:2009aa}. In addition, if the dark scalar sector undergoes a strong first–order phase transition, it can source a stochastic gravitational–wave background within reach of planned space–based interferometers \cite{Caprini:2019egz,Amaro-Seoane:2017vxo}. These avenues are complementary to laboratory searches and remain informative even when direct and indirect detection are intrinsically suppressed.

\emph{Indirect detection}
If Sommerfeld enhancement or bound-state formation are active, the present-day annihilation rate can be boosted well above the perturbative expectation, with a characteristic \emph{velocity dependence} that makes low-dispersion targets especially promising. In the Milky Way and dwarf spheroidal galaxies ($v\sim10^{-3}$ and $10^{-4}$, respectively), Sommerfeld enhancement can raise $\langle \sigma v_{\rm{rel}} \rangle$ by factors $S\!\sim\!2\pi\alpha_{\rm eff}/v$ in the Coulomb limit or up to a saturated plateau $S_{\max}\!\sim\!\alpha_{\rm eff} m_\chi/m_\phi$ for a Yukawa force; near zero-energy levels, narrow resonances can further amplify the rate, though still bounded by $S_{\max}$ \cite{Cassel:2009wt,Feng:2010zp}. For secluded models with a dark photon $A'$ that promptly decays to SM fermions, the resulting signatures include: (i) \emph{gamma rays} from hadronic final states (via $\pi^0\!\to\!\gamma\gamma$) and from inverse-Compton upscattering of Sommerfeld-enhancement-enhanced $e^\pm$; (ii) \emph{cosmic-ray} $e^\pm$ and $\bar{p}$ fluxes; and (iii) \emph{radio/synchrotron} emission from SE-boosted $e^\pm$ in magnetic fields (see \cite{Essig:2013lka} for overviews). Bound-state formation can further increase late-time depletion and may add distinctive channels: radiative capture scales roughly as $\sigma_{\rm BSF} v \propto \alpha_{\rm eff}^5/(m_\chi^2 v)$ in the Coulombic limit, enhancing the signal at low $v$; capture or de-excitation typically emits a dark photon that subsequently decays to SM states, injecting additional $e^\pm/\gamma$ power \cite{An:2016gad,Mitridate:2017izz}. Because Sommerfeld enhancement and bound-state formation also operate at recombination-era velocities ($v\!\sim\!10^{-8}$) until Yukawa saturation $S_{\rm{max}}$, \emph{CMB energy-injection} bounds can become very constraining for $s$-wave channels (effectively limiting today’s boosted $\langle \sigma v_{\rm{rel}} \rangle$), whereas long-lived or invisibly decaying mediators can shift power to later times and weaken standard gamma-ray constraints while strengthening cosmological/BBN probes \cite{An:2016gad,Mitridate:2017izz}.
If Sommerfeld enhancement or bound-state formation are active, the present-day annihilation rate can be boosted well above the perturbative expectation, with a characteristic \emph{velocity dependence} that makes low-dispersion targets especially promising. In the Milky Way and dwarf spheroidal galaxies ($v\sim10^{-3}$ and $10^{-4}$, respectively), Sommerfeld enhancement can raise $\langle \sigma v_{\rm{rel}} \rangle$ by factors $S\!\sim\!2\pi\alpha_{\rm eff}/v$ in the Coulomb limit or up to a saturated plateau $S_{\max}\!\sim\!\alpha_{\rm eff} m_\chi/m_\phi$ for a Yukawa force; near zero-energy levels, narrow resonances can further amplify the rate, though still bounded by $S_{\max}$ \cite{Cassel:2009wt,Feng:2010zp}. For secluded models with a dark photon $A'$ that promptly decays to SM fermions, the resulting signatures include: (i) \emph{gamma rays} from hadronic final states (via $\pi^0\!\to\!\gamma\gamma$) and from inverse-Compton upscattering of Sommerfeld-enhancement-enhanced $e^\pm$; (ii) \emph{cosmic-ray} $e^\pm$ and $\bar{p}$ fluxes; and (iii) \emph{radio/synchrotron} emission from SE-boosted $e^\pm$ in magnetic fields (see \cite{Essig:2013lka} for overviews). Bound-state formation can further increase late-time depletion and may add distinctive channels: radiative capture scales roughly as $\sigma_{\rm BSF} v \propto \alpha_{\rm eff}^5/(m_\chi^2 v)$ in the Coulombic limit, enhancing the signal at low $v$; capture or de-excitation typically emits a dark photon that subsequently decays to SM states, injecting additional $e^\pm/\gamma$ power \cite{An:2016gad,Mitridate:2017izz}. Because Sommerfeld enhancement and bound-state formation also operate at recombination-era velocities ($v\!\sim\!10^{-8}$) until Yukawa saturation $S_{\rm{max}}$, \emph{CMB energy-injection} bounds can become very constraining for $s$-wave channels (effectively limiting today’s boosted $\langle \sigma v_{\rm{rel}} \rangle$), whereas long-lived or invisibly decaying mediators can shift power to later times and weaken standard gamma-ray constraints while strengthening cosmological/BBN probes \cite{An:2016gad,Mitridate:2017izz}.

\subsection{CMB constraints and spectral distortions}
\label{sec:cmb}

Energy injected into the primordial plasma after recombination modifies the ionization history and damps CMB temperature and polarization anisotropies. In the standard formalism this effect is captured by the parameter $p_{\rm ann}\!\equiv\!f_{\rm eff}\,\langle \sigma v_{\rm{rel}} \rangle/m_\chi$, where $f_{\rm eff}$ encodes the energy–deposition efficiency \cite{Slatyer:2015jla}. Planck bounds require $p_{\rm ann}\lesssim\,$few$\times10^{-28}\,{\rm cm^3\,s^{-1}\,GeV^{-1}}$ \cite{Aghanim:2018eyx}. In secluded models this limit is often naturally weak: (i) many dominant channels are $p$–wave or otherwise velocity suppressed, so the rate at recombination ($v\!\sim\!10^{-8}$) is negligible; (ii) when $s$–wave is present it is typically off resonance today because the ultra–narrow mediator (tiny portal) makes it unlikely that $2m_\chi$ sits on the pole; (iii) cascades $\chi\chi\!\to\!\phi\phi\!\to\!\mathrm{SM}$ soften the spectra and reduce $f_{\rm eff}$ \cite{Slatyer:2015jla}. For our benchmarks, CMB anisotropy limits are therefore typically less constraining than collider and intensity–frontier searches.

Long–lived mediators can also affect the CMB if they decay during or after recombination. In our benchmarks the portal couplings are chosen large enough that $\phi$ decays well before BBN, $\tau_\phi\!\ll\!1$\,s (Sec.~\ref{sec:BBN}), i.e.\ many orders of magnitude earlier than recombination ($t\!\sim\!4\times10^{5}$\,yr). Consequently, mediator decays do not distort the recombination history. Only for extremely feeble portals, yielding $\tau_\phi\!\gtrsim\!10^{12}\!-\!10^{13}$\,s, would CMB anisotropies and the ionization history be impacted; such parameters are outside the regimes considered here and are tightly constrained by cosmology on their own.

CMB $\mu$/$y$ spectral distortions arise from electromagnetic energy injection at $10^{4}\!\lesssim\! z \!\lesssim\! 2\times10^{6}$ \cite{Chluba:2011hw,Chluba:2013wsa}. In our benchmarks the mediator decays long before this epoch, $\tau_\phi\!\ll\!1$\,s ($z\!\gg\!10^{6}$), so its decay products thermalize and generate no distortions. Residual annihilation during the distortion era is also subdominant for the reasons above. Consequently, COBE/FIRAS limits ($|\mu|,|y|\!\lesssim\!10^{-5}$) and projected PIXIE/PRISM sensitivities ($|\mu|\!\sim\!10^{-8}$–$10^{-9}$) \cite{Kogut:2011xw,Andre:2013nfa} are not expected to constrain our parameter space. By contrast, in ultra–feeble–portal scenarios with lifetimes approaching the $\mu/y$ era, spectral–distortion probes would become powerful and complementary to collider searches.

\section{Conclusions}
\label{sec:conclusion}

We presented a comprehensive analysis of the \emph{secluded} DM scenario originally presented in Refs.~\cite{Pospelov:2007mp,Pospelov:2008jd}, in which the processes that set the relic abundance are decoupled from those that control the spin–independent nuclear cross section and indirect detection annihilation cross section. The motivation is clear: present direct– and indirect-detection limits stringently constrain models where the same portal governs both early–Universe annihilation and present–day scattering.
In particular, in secluded models with DM mass $m_\chi$ above a lighter mediator $\phi$ (or a Higgs–mixed scalar that plays the role of a portal), the observed relic density is predominantly achieved through $\chi\bar\chi \to \phi \phi$ within the hidden sector. The portal to the SM can thus be highly suppressed, naturally evading direct- and indirect-detection bounds.

%We showed that this mechanism reproduces the measured relic abundance across broad regions of parameter space while keeping the portal small enough to suppress nuclear scattering and annihilation cross sections into SM particles. 
We examined three minimal realizations:
\begin{itemize}
\item \textbf{Model I ($U(1)_X$ with dark photon $A'$).} A new gauge boson $A'$ couples to the DM particle $\chi$ and mixes kinetically with SM hypercharge boson via a portal coupling $\epsilon$. In the secluded regime ($m_\chi>m_{A'}$), the relic density is set by $\chi\bar\chi\to A'A'$, while spin-independent scattering and indirect detection cross section scales as $\propto \epsilon^2$ and can be made negligible.
\item \textbf{Model II (scalar DM with scalar mediator $K$).} A scalar DM $S$ annihilates dominantly into pairs of a light scalar mediator $K$ through quartic/cubic interactions in the dark sector. The mediator’s Yukawa–like coupling to SM fermions controls direct and indirect detection and can be taken small without affecting freeze–out.
\item \textbf{Model III (Dirac DM with Higgs–mixed singlet $H_p$).} A singlet scalar $H_p$ mixes with the SM Higgs by an angle $\alpha$ and couples to Dirac DM via a Yukawa $y_p$. For $m_\chi>m_{H_p}$ and small mixing, $\chi\bar\chi\to H_pH_p$ sets the relic density, while spin-independent scattering is suppressed by $\tan{2\alpha}$.
\end{itemize}

Across all three models, choosing portal parameters below $\mathcal{O}(10^{-3})$ renders direct– and indirect-detection constraints weak, while the relic density is fixed by hidden–sector annihilations rather than by DM annihilating into SM particles.
Despite the small portal couplings, the mediators remain in chemical equilibrium with the SM bath at freeze–out and the mediator decay into SM fermions happens well before BBN.

Even with next–generation detectors such as DARWIN \cite{DARWIN:2016hyl} approaching the neutrino floor, large regions of the secluded parameter space will remain viable. Secluded DM thus offers robust, economical, and thermal scenarios testable at collider experiments that evade current and future direct–detection limits while remaining consistent with cosmology.

\begin{acknowledgments} 
M.D.M. acknowledges support from the research grant {\sl TAsP (Theoretical Astroparticle Physics)} funded by Istituto Nazionale di Fisica Nucleare (INFN) and from the Italian Ministry of University and Research (MUR), PRIN 2022 ``EXSKALIBUR – Euclid-Cross-SKA: Likelihood Inference Building for Universe’s Research'', Grant No. 20222BBYB9, CUP I53D23000610 0006, and from the European Union -- Next Generation EU.

On a personal note, M.D.M. would like to thank his wife, Chiara, for reminding him that dark matter probably does not exist. 
This remark inspired the work on the secluded dark matter scenario presented here — a model that still does not satisfy her skepticism, but might at least explain why we have not detected it yet.
\end{acknowledgments}

\bibliographystyle{unsrt}
\bibliography{refs}

\newpage

\end{document}